\documentclass[fleqn,usenatbib]{mnras}

\usepackage{newtxtext,newtxmath}

\usepackage[T1]{fontenc}

\DeclareRobustCommand{\VAN}[3]{#2}
\let\VANthebibliography\thebibliography
\def\thebibliography{\DeclareRobustCommand{\VAN}[3]{##3}\VANthebibliography}

\usepackage{graphicx}	
\usepackage{amsmath}	
\usepackage{tikz}
\usepackage{tikz-3dplot}
\usepackage{caption}
\usepackage{subcaption}
\usepackage[skip=0.5ex]{subcaption}
\usepackage{xcolor}
\usepackage{comment}

\newcommand{\frank}[0]{\textsc{frank}}
\newcommand{\un}[1]{\mathbfit{#1}}
\newcommand{\unun}[1]{\mathbfss{#1}}
\newcommand{\mean}{\boldsymbol{\mu}}

\newcommand{\seba}[1]{\textcolor{teal}{\textbf{Seba:} #1}}

\title[Deprojecting exoKuiper belts]{Deprojecting and constraining the vertical thickness of exoKuiper belts}

\author[James Terrill et al.]{James Terrill$^{1}$, Sebastian Marino$^{1,2,3}$\thanks{E-mail:sebastian.marino.estay@gmail.com}, Richard A. Booth$^{4,5}$, Yinuo Han$^{1}$, Jeff Jennings$^{6}$ and Mark C. Wyatt$^{1}$.\\
  $^{1}$Institute of Astronomy, University of Cambridge, Madingley Road, Cambridge CB3 0HA, UK\\
  $^{2}$Jesus College, University of Cambridge, Jesus Lane, Cambridge CB5 8BL, UK\\
  $^{3}$Department of Physics and Astronomy, University of Exeter, Stocker Road, Exeter, EX4 4QL, UK\\
  $^{4}$Astrophysics Group, Imperial College London, Blackett Laboratory, Prince Consort Road, London SW7 2AZ, UK\\
  $^{5}$School of Physics and Astronomy, University of Leeds, Leeds, LS2 9JT, UK\\
  $^{6}$Department of Astronomy and Astrophysics, 525 Davey Laboratory, The Pennsylvania State University, University Park, PA 16802, USA\\  
}

\pubyear{2033}

\begin{document}
\label{firstpage}
\pagerange{\pageref{firstpage}--\pageref{lastpage}}
\maketitle

\begin{abstract}
Constraining the vertical and radial structure of debris discs is crucial to understanding their formation, evolution and dynamics. To measure both the radial and vertical structure, a disc must be sufficiently inclined. However, if a disc is too close to edge-on, deprojecting its emission becomes non-trivial. In this paper we show how \textsc{Frankenstein}, a non-parametric tool to extract the radial brightness profile of circumstellar discs, can be used to deproject their emission at any inclination as long as they are optically thin and axisymmetric. Furthermore, we extend \textsc{Frankenstein} to account for the vertical thickness of an optically thin disc ($H(r)$) and show how it can be constrained by sampling its posterior probability distribution and assuming a functional form (e.g. constant $h=H/r$), while fitting the radial profile non-parametrically. We use this new method to determine the radial and vertical structure of 16 highly inclined debris discs observed by ALMA. We find a wide range of vertical aspect ratios, $h$, ranging from $0.020\pm0.002$ (AU Mic) to $0.20\pm0.03$ (HD~110058), which are consistent with parametric models. We find a tentative correlation between $h$ and the disc fractional width, as expected if wide discs were more stirred. Assuming discs are self-stirred, the thinnest discs would require the presence of at least 500~km-sized planetesimals. The thickest discs would likely require the presence of planets. We also recover previously inferred and new radial structures, including a potential gap in the radial distribution of HD~61005. Finally, our new extension of \textsc{Frankenstein} also allows constraining how $h$ varies as a function of radius, which we test on 49~Ceti, finding that $h$ is consistent with being constant.




\end{abstract}

\begin{keywords}
circumstellar matter -- planetary systems -- methods: numerical -- techniques: interferometric -- planets and satellites: dynamical evolution and stability
\end{keywords}



\section{Introduction}\label{sec:intro}

Debris discs are a ubiquitous component of planetary systems, analogues of the Asteroid and Kuiper belts, and readily found around 20\% of nearby AFGK stars \citep[e.g.][]{Su2006, Eiroa2013, Sibthorpe2018}. These  discs are made of solids in a wide size distribution from km-sized planetesimals down to $\mu$m-sized dust grains. This size distribution is maintained by a collisional cascade that grinds solids down to sizes small enough to be blown-out by radiation pressure or stellar winds \citep{Wyatt2008, Hughes2018, Marino2022debris_disc_book}. Kuiper belt analogues (or exoKuiper belts), in particular, can be orders of magnitude brighter than planets at tens of au and thus provide a unique window to study the formation and dynamics of planetary systems. To this end, ALMA has been fundamental to constraining the distribution of large grains, for which radiation forces are negligible, and thus trace better the dynamics and location of planetesimals. Due to its high sensitivity and variable resolution, ALMA observations have provided precise measurements of the structure of debris discs.

The observed structure of debris discs provides important clues to the properties of hypothetical embedded planets and give insight into their dynamics. Radial structure can be used to infer the presence of inner planets truncating the disc  \citep[e.g.][]{Quillen2006, Chiang2009, Mustill2012, Nesvold2015}, embedded planets clearing gaps \citep[e.g.][]{Marino2018hd107, Marino2019, MacGregor2018, Marino2020hd206, Nederlander2021} whose widths can constrain the planet masses and migration histories \citep[e.g.][]{Morrison2015, Friebe2022}, and the level of dynamical stirring \citep{Marino2021}. Non-axisymmetric structures such as clumps and disc eccentricities can also reveal the dynamical shaping by planets \citep[e.g.][]{Kalas2005, Wyatt2006, Dent2014, Faramaz2019}. Finally, high-resolution ALMA observations have started to constrain the vertical thickness of a few debris discs, typically revealing vertical aspect ratios of ${\sim}2-20\%$ \citep{Kennedy2018, Matra2019betapic, Daley2019, Marino2019, Marino2021, Hales2022, Marshall2023}. Such measurements directly probe the distribution of orbital inclinations, and thus can be used to constrain the mass of the bodies stirring the disc. Moreover, measurements of the vertical thickness at different wavelengths may also constrain the internal strength of solids \citep{Vizgan2022}.

Despite the progress described above, it has been challenging to determine both the radial and vertical structure of systems. This is because the radial structure is best studied in face-on systems, while the vertical structure is only accessible for highly inclined discs, which then obscures the radial structure. Parametric models have been used to fit the data and derive basic radial and vertical properties \citep[e.g.][]{Marino2016, Marino2019, Kennedy2018, Matra2019betapic}, but such methods rely on assuming parametric models that could bias such estimates. Very recently, \cite{Han2022} developed a tool called \textsc{Rave} that can deproject the emission of edge-on discs non-parametrically and constrain their vertical thickness using thermal emission images. Whilst \textsc{Rave} has been demonstrated to work well with images of edge-on discs, there has not been a method that could work directly with the interferometric visibilities measured by ALMA (making full use of its resolution power) and with discs that are not edge-on. 

In this paper we present a new approach to simultaneously deproject the emission of debris discs and constrain their vertical structure, independent of their observed inclination\footnote{Although this method is applicable to any inclination, the constraints on the vertical structure depend on the resolution (uv-coverage) and inclination of a disc.}. In order to do so, we develop a new extension of \textsc{Frankenstein} \citep[][\textsc{Frank} hereafter]{Jennings2020}. \textsc{Frank} non-parametrically fits the real component of the azimuthally averaged visibilities to obtain a 1D radial brightness profile for a disc. The base version of the code assumes the disc is flat, while our new extension takes into account the vertical thickness of optically thin emission, which can be fitted in an iterative method.

This paper is structured as follows. In \S\ref{sec:bkg} we introduce the key definitions to describe the emission of a debris disc and its visibilities. In \S\ref{sec:newfrank} we show how the vertical thickness of debris discs affects the visibilities and can be incorporated into   \textsc{frank}. In \S\ref{sec:tests} we test the new algorithm on simulated data and constrain its accuracy. In \S\ref{sec:fitdata} we apply our new extension of \textsc{frank} to archival ALMA data. Finally, in \S\ref{sec:dis} and \S\ref{sec:conclusions} we discuss our findings and summarise our conclusions.

\section{Background and definitions}
\label{sec:bkg}
In this section we introduce a few key concepts to describe the structure of debris discs, their on-sky emission, and the measurement of this emission with interferometers such as ALMA.

\subsection{Surface density, aspect ratio, and emissivity}\label{subsec:defn_of_h}

We start by defining the disc properties in cylindrical coordinates $(r, z, \phi)$, with the origin at the central star position and the disc midplane lying at $z=0$. We will assume discs are axisymmetric and thus their mass density ($\rho$) is only a function of $r$ and $z$. Vertically, the density is assumed to follow a Gaussian distribution 
\begin{equation}
    \rho(r,z) = \Sigma(r) \frac{\exp\left(-\frac{z^2}{2H(r)^2}\right)}{\sqrt{2\pi} H(r)},\label{eqn:disc_density}
\end{equation}
where $\Sigma(r)$ is the surface density and $H(r)$ the vertical standard deviation or scale height. We will refer to the ratio $h=H/r$ as the disc aspect ratio. The aspect ratio is directly related to the dispersion of orbital inclinations in the disc with $h=i_{\rm rms}/\sqrt{2}$ \citep[]{Matra2019betapic}. Finally, since debris discs are optically thin and vertically thin ($H\ll r$), their equilibrium temperature will scale approximately as $\frac{1}{\sqrt{r}}$. 

Unless the vertical distribution ($H$) varies significantly as a function of grain size, the volume emissivity at long wavelengths will also have an approximately Gaussian structure (this assumption is discussed in \S\ref{sec:gaussian}),
\begin{align}
    \epsilon_\nu(r, z) &= \kappa_\nu(r) B_\nu[T(r)] \Sigma(r)  \frac{\exp\left(-\frac{z^2}{2H(r)^2}\right)}{\sqrt{2\pi}H(r)} \\ 
    &= I_\nu (r) \frac{\exp\left(-\frac{z^2}{2H(r)^2}\right)}{\sqrt{2\pi}H(r)}. \label{eqn:emiss}
\end{align}
Here $\kappa_\nu(r)$ and $B_\nu [T(r)]$ are the opacity and Planck function. In the final equality we have introduced $I_\nu (r) = \kappa_\nu(r) B_\nu[T(r)] \Sigma(r)$. For any disc that is optically thin $I_\nu(r)$ is simply the surface brightness of a face-on disc. 

For a disc that is not exactly face on, individual lines of sight include contributions from a range of radii. Maintaining our optically thin assumption, we may write the sky brightness at a point $(x,y)$ as
\begin{equation}
    I_{\rm s}(x, y) = \int \epsilon'_\nu(x, y, z) {\rm d}z,
\end{equation}
where $\epsilon'_\nu(x, y, z)$ is determined from the face-on (or de-projected) emissivity, $\epsilon_\nu(x_d, y_d, z_d)$, by a rotation of the coordinate system. Without loss of generality, we may consider rotations about the $x$-axis only (i.e. a disc with a position angle of $90^\circ$)\footnote{For real data, we account for the position angle of the disc major axis by rotating the data in visibility space. In this work we have assumed the position angle to be well-known, which is true for the studied sample in \S\ref{sec:fitdata}.}, as shown in \autoref{fig:inc_defn_diag}. Thus, $x_d = x$, $y_d = y \cos i - z \sin i$, and $z_d = y \sin i + z \cos i$ (where $i$ is the inclination). 

In the next section we will discuss how the vertical structure can be taken into account directly in Fourier space, but first we consider the special case of a razor-thin disc ($H\rightarrow0$), for which we arrive at the well-known result $I_{\rm s}(x, y) = I_\nu(x_d, y_d) / \cos i$. By $I_\nu(x_d, y_d)$ we explicitly mean $I_\nu(r)$ where $r^2 = x_d^2 + y_d^2$ and $y_d =y / \cos i$ (since $z_d = 0$).
It should be noted that the $I_{\rm s}(x,y) \propto 1/\cos i$ scaling only applies for optically thin emission; for an optically thick disc the $1/\cos i$ term is absent. 

\tdplotsetmaincoords{60}{110}
\pgfmathsetmacro{\rvec}{1}
\pgfmathsetmacro{\thetavec}{30}
\pgfmathsetmacro{\phivec}{90}

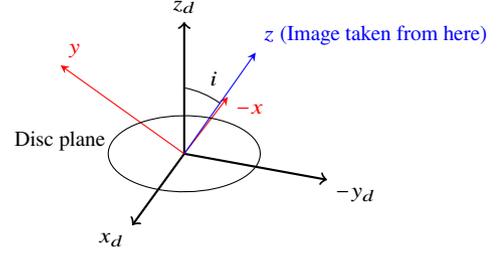
\begin{figure}
\begin{center}
\begin{tikzpicture}[scale=2,tdplot_main_coords]
\coordinate (O) at (0,0,0);
\draw[thick,->] (0,0,0) -- (1,0,0) node[anchor=north east]{$x_d$};
\draw[thick,->] (0,0,0) -- (0,1,0) node[anchor=north west]{$-y_d$};
\draw[thick,->] (0,0,0) -- (0,0,1) node[anchor=south]{$z_d$};
\tdplotsetcoord{yp}{\rvec}{-2*\thetavec}{\phivec}
\draw[-stealth,color=red] (O) -- (yp) node[above right]{$y$};
\tdplotsetcoord{xp}{0.8*\rvec}{3*\thetavec}{179}
\draw[-stealth,color=red] (O) -- (xp) node[below right]{$-x$};
\tdplotsetcoord{P}{\rvec}{\thetavec}{\phivec}
\draw[-stealth,color=blue] (O) -- (P) node[above right]{$z$ (Image taken from here)};
\tdplotsetthetaplanecoords{\phivec}
\tdplotdrawarc[tdplot_rotated_coords]{(0,0,0)}{0.5}{0}%
    {\thetavec}{anchor=south west}{$i$}
\tdplotdrawarc[tdplot_main_coords]{(0,0,0)}{0.5}{449}%
    {\phivec}{anchor= east}{Disc plane}
\end{tikzpicture}
\end{center}
\caption{Illustration of the sky-projected coordinate system $(x,y)$ and disc or deprojected coordinates $(x_d, y_d, z_d)$.}
\label{fig:inc_defn_diag}
\end{figure}

\subsection{Visibility}
\label{subsec:vis}

An interferometer such as ALMA samples the sky brightness as an ensemble of complex visibilities,
\begin{equation}
    V_s(u,v)=\iint_S PB(x,y) I_s(x,y) \exp{\left[-2\pi i (ux+vy)\right]}dx\,dy ,\label{eqn:vis_defn_1}
\end{equation}
where $S$ is the region of the sky over which the integral is taken and $PB(x,y)$ is the antenna primary beam, with a peak of 1 and FWHM ${\sim}\lambda/12$~m for ALMA's 12m diameter antennas. In this paper we will focus on discs smaller than the primary beam and for which we will approximate $PB(x,y)\approx 1$. Now $V_s(u,v)$ is simply the Fourier Transform of the sky brightness.

For the razor-thin disc,
\begin{equation}
    V_s(u,v)=\iint_S \frac{I_\nu(x, y/\cos i)}{\cos i} \exp{\left[-2\pi i (ux+vy)\right]}dx\,dy.
\end{equation}
If we denote $V_0(u, v)$ as the visibility distribution we would observe if the disc were face-on, then by changing variables to $y_d = y/\cos i$ it is straightforward to demonstrate that $V_s(u, v) = V_0(u, v \cos i)$. This means that we can de-project in Fourier space via a simple transformation of the u-v coordinates to $(u_d, v_d) = (u, v \cos i)$.  \footnote{Again it should be noted that for an optically thick disc $V_s(u, v) \propto \cos i$. By default \frank{} assumes optically thick emission and corrects for this proportionality for its standard (razor thin) disc model. This is usually the desired behaviour since it results in an extracted radial profile that would have the same flux as one extracted from an image plane analysis (in the limit of infinite resolution). Such an assumption is however not valid if one assumes an optically thin disc, which is the case for the debris disc model developed here. The new version of \frank{} includes now an option to consider optically thin or thick emission.} As a disc becomes more inclined and its emission appears compressed along the $y$-axis in the image space, the morphology of its visibilities becomes stretched in the $v$-axis on the Fourier space.

For a disc with a finite thickness we can use these ideas to compute the visibilities without needing to directly compute the sky brightness, as long as we assume the emission is optically thin. From \autoref{eqn:emiss} and \autoref{eqn:vis_defn_1}, we have
\begin{equation}
    V_s(u,v)=\iiint_V \epsilon'_\nu(x,y,z) \exp{\left[-2\pi i (ux+vy)\right]}dx\,dy\,dz
\end{equation}
Substituting the de-projected coordinates ($x_d,y_d,z_d$) for the sky-plane coordinates ($x=x_d$, $y=y_d\cos i + z_d \sin i$, $z = -y_d \sin i + z_d \cos i$) and noting that $dx\,dy\,dz=dx_d\,dy_d\,dz_d$, we arrive at 
\begin{align}
    V_s(u,v)=\iiint_V & I_\nu(r)  \frac{\exp\left(-\frac{z_d^2}{2H(r)^2}\right)}{\sqrt{2\pi}H(r)} \nonumber \\
    & \times \, \exp{\left[-2\pi i (u_d x_d + v_dy_d + w_d z_d)\right]}   dx_d\,dy_d\,dz_d,
\end{align}
where $r^2 = x_d^2 + y_d^2$ and $(u_d, v_d, w_d) = (u, v \cos i , v \sin i)$. Completing the Fourier Transform in the $z_d$-direction we arrive at:
\begin{align}
    V_s(u,v)=\iint_S & I_\nu(r) \exp\left\{-\frac{1}{2}[2\pi w_d H(r)]^2\right\} \nonumber \\
    & \times \,  \exp{\left[-2\pi i (u_d x_d + v_dy_d)\right]} dx_d\,dy_d.
    \label{eqn:vis_thick}
\end{align}
It is straightforward to confirm that this expression reproduces our previous expressions for inclined, optically-, and razor-thin discs when $H(r) = 0$. Hence for discs that are sufficiently thin (small $H$) or close to face-on (small $w_d$) the visibilities are hardly modified by the vertical structure, and therefore any tool that can infer brightness profile from the visibilities, such as \frank{}, can also be used for inclined discs. 

Although we cannot proceed any further analytically unless $H(r)$ and $I_\nu(r)$ take a particularly simple form, we will show that \frank{} can easily be modified to incorporate the exponential term arising from the non-zero thickness. This allows \frank{} to be used for highly inclined discs if they are optically thin. Before doing so, we briefly review the standard approach used to infer $I_\nu(r)$ in \frank{}.

\section{A vertical extension to Frankenstein}
\label{sec:newfrank}

\frank{} reconstructs a disc’s radial intensity profile by assuming azimuthal symmetry and non-parametrically fitting the real component of the 
deprojected visibilities in 1D \citep{Jennings2020}. The model obtains super-resolution to recover disc features 
that are under-resolved in a standard \texttt{CLEAN} image. \frank{} has been applied to tens of protoplanetary discs \citep[e.g.,][]{Jennings2022} and a few debris discs \citep[Imaz-Blanco et al. submitted]{Marino2020hd206}, revealing new radial features.
In comparison to parametric radiative transfer models with multiple parameters that can be fit to ALMA data using MCMC methods \citep[e.g.][]{Marino2016} over hours on multiple CPUs, \textsc{frank} performs a fit in $\lesssim 1$ minute on a single CPU.

\subsection{Frankenstein applied to razor-thin discs}
\label{subsec:frank_applied_to_flat}

\textsc{Frank} infers the intensity profile, $I(r)$, by using a Discrete Hankel Transform to map the intensities at a set of radial locations, $r_k$, to the visibility space. The intensities, $I(r_k)$, are then inferred by fitting the observed visibilities and regularized using a Gaussian process. Below we briefly describe the most important equations to understand the method, to later expand it to consider the vertical thickness of discs.

We start by recalling that for an axisymmetric disc (after deprojection), the 2D Fourier transform reduces to 1D as the Hankel transformation with Bessel function kernels \citep{Bracewell2000, Thompson2017}
\begin{eqnarray}
    V_s(q)=\int_0^{R_{\textrm{out}}} I_s(r)J_0(2\pi q r)2\pi r dr, \label{eqn:v_to_i_defn_pt1} \\
    I_s(r)=\int_0^{Q_{\textrm{max}}} V_s(q)J_0(2\pi q r)2\pi qdq,\label{eqn:v_to_i_defn}
\end{eqnarray}
where $q=\sqrt{u_d^2+v_d^2}$ and $r=\sqrt{x_d^2+y_d^2}$. Assuming $I(r)=0$ beyond some radial distance $R_{\rm out}$ and $V(q)=0$ beyond some baseline $Q_{\max}$ we can expand $V(q)$ and $I(r)$ in a Fourier-Bessel series 
\begin{eqnarray}
        I_s(r)=\sum_{k=1}^\infty\alpha_kJ_0\left(\frac{j_{0k}r}{R_{\textrm{out}}}\right),\label{eqn:i_bessel_series} \\
        V_s(q)=\sum_{k=1}^\infty\beta_kJ_0\left(\frac{j_{0k}q}{Q_{\textrm{max}}}\right), \label{eqn:Vis_pred}
\end{eqnarray}
where $j_{0,k}$ is the $k$th zero of $J_0$, and the coefficients $\alpha_k$ and $\beta_k$ can be computed as
\begin{eqnarray}
    \alpha_k=\frac{1}{\pi R^2_{\textrm{out}}J^2_1\left(j_{0k}\right)}V_s\left(\frac{j_{0k}}{2\pi R_{\textrm{out}}}\right), \\
    \beta_k=\frac{1}{\pi Q^2_{\textrm{max}}J^2_1\left(j_{0k}\right)}I_s\left(\frac{j_{0k}}{2\pi Q_{\textrm{max}}}\right).
\end{eqnarray}

In practice, this infinite series must be truncated after $N$ terms, and therefore $I_s$ is determined by baselines below $q = \frac{j_{0, N+1}}{2\pi R_{\textrm{out}}}$, and $V_s$ is determined by radii smaller than $r=\frac{j_{0,N+1}}{2\pi Q_{\textrm{max}}}$. We enforce $Q_{\textrm{max}}=j_{0,N+1}/2\pi R_{\textrm{out}}$, for the DHT, with the collocation points:
\begin{equation}
    r_k\coloneqq R_{\textrm{out}}j_{0k}/j_{0,N+1}.
\end{equation}
\begin{equation}
    q_k\coloneqq j_{0k}/2\pi R_{\textrm{out}}.
\end{equation}
Now the intensity is a vector $\un{I}_s$, with components $I_k\coloneqq I_s(r_k)$. For a given set of intensities, $\un{I}_s$, \frank{} uses \autoref{eqn:Vis_pred} to compute the `model visibilities', 
\begin{equation}
    \un{V}_s(\un{q}) = \unun{H}(\un{q}) \un{I}_s \label{eqn:model_vis}
\end{equation}
where 
\begin{equation}
    \unun{H}(\un{q})_{jk}=\frac{4\pi R^2_{\textrm{out}}}{j^2_{0,N+1}J_1^2(j_{0k})}J_0\left(2\pi q_j R_{\textrm{out}}\frac{j_{0k}}{j_{0,N+1}}\right).\label{eqn:defining_Hjk}
\end{equation}

For a set of measured visibilies, $\un{V}$, with corresponding baselines $\un{q}$ and statistical weights, $\un{w}$, the intensity is inferred from the posterior probability distribution,
\begin{align}
P(\un{I}|\un{V}, \un{p}) &= \frac{\mathcal{G}\left(\un{V} - \unun{H}(\un{q}) \un{I}_s, \unun{N}\right) \mathcal{G}\left(\un{I}_s, \unun{S}(\un{p})\right)}{P(\un{V}|\un{p})}, \label{eqn:likelihood}
\end{align}
where $\mathcal{G}(\un{x}, \mathbf{\Sigma})$ is a Gaussian distribution with mean zero and covariance $\mathbf{\Sigma}$, $\unun{N}=\textrm{diag}(1/\un{w})$, and $\unun{S}(\un{p})$ is the covariance of the Gaussian process prior. For details of this prior and the parameters, $\un{p}$, upon which it depends, see \citet{Jennings2020}. Following \citet{Jennings2020} we will refer to $\un{p}$ as the power spectrum parameters. Since we have not modified either the prior or the way the parameters are determined, we do not repeat the description here.

Since $P(\un{I}|\un{V}, \un{p})$ is the product of two Gaussians, it is also a Gaussian, and has covariance $\unun{D}$ and mean $\mean$, 
\begin{align}
  \unun{D} &= \left( \unun{M} + \unun{S}(\un{p})^{-1} \right)^{-1}, \label{eqn:D} \\
   \mean &= \unun{D} \,\un{j}, \label{eqn:defining_trans}
\end{align}
where
\begin{align}
  \unun{M} &= \unun{H}(\un{q})^T \unun{N}^{-1} \unun{H}(\un{q}), \\
  \un{j} &=\unun{H}^T\unun{N}^{-1}\un{V}. \label{eqn:j}
\end{align}
Finally, $\un{I}_s = \mean$ is used as the inferred brightness.

\subsection{Treating vertical thickness in \frank{}}

In \S\ref{subsec:vis} we showed that visibilities of a disc with a Gaussian vertical structure are given by \autoref{eqn:vis_thick}. Assuming both $I(r)$ and $H(r)$ are axisymmetric, then we obtain a Hankel Transform in which $I_s(r)$ is replaced with $I_s(r) \exp\left\{-\frac{1}{2}[2\pi w_d H(r)]^2\right\}$. Equation~\ref{eqn:v_to_i_defn_pt1} then becomes
\begin{equation}
    V_s(q, w_d)=\int_0^{R_{\textrm{out}}} I_s(r) \exp\left\{-\frac{1}{2}[2\pi w_d H(r)]^2\right\} J_0(2\pi q r)2\pi r dr. \\    
\end{equation}
This implies that the Fourier-Bessel series can also be modified to account for the vertical structure by making the substitution
\begin{equation}
    \beta_k \rightarrow \beta'_k(w_d) = \beta_k \times \exp\left\{-\frac{1}{2}[2\pi w_d H(r_k)]^2\right\}. 
\end{equation}
It follows from this that, if $H(r)$ is a known function, we can infer $I(r)$ directly from the visibilities by modifying the mapping between the intensity at the collocation points $\un{I}(r_k)$ and the model visibilities $\un{V}_s(q)$. We make the substitution
\begin{equation}
    \unun{H}(\un{q})_{jk}\xrightarrow[]{}\unun{H}(\un{q})_{jk}\exp\left(-\frac{1}{2}\left[2\pi w_dH(r_j)\right]^2\right).
\end{equation}
This change to $\unun{H}(\un{q})$ is used in Equations~\ref{eqn:D}--\ref{eqn:j} or when computing the model visibilities (\autoref{eqn:model_vis}),
with the code otherwise unchanged. To be explicit, the quantity inferred by this module in \frank{} is the vertically integrated intensity of a face-on disk, i.e. $I_\nu(r)$.

\subsection{Frankenstein - parameters}\label{subsec:frank_params}

The algorithm has five input parameters (in addition to a supplied or internally determined disc geometry), two of which alter the model's Gaussian process prior and should thus be varied to explore the significance of features recovered in the radial brightness profile. These are:
\begin{itemize}
    \item $\alpha$ effectively sets the signal-to-noise threshold at which the model no longer attempts to fit the visibilities. In practice this sets a maximum baseline out to which the data are fit. The range recommended by \cite{Jennings2020} is $1.0 - 1.3$.
    \item $w_{\textrm{smooth}}$ is a parameter included to counteract underestimated uncertainties arising from incomplete $(u, v)$ sampling by encouraging smoothness in the power spectrum parameters, $\un{p}$. It has little effect on the reconstructed brightness profile. The range recommended by \cite{Jennings2020} is $10^{-4} - 10^{-1}$, with higher values having smoother $\un{p}$.
\end{itemize}
Our new, vertically aware extension of \textsc{frank} introduces an additional parameter:
\begin{itemize}
    \item Our extension allows for the vertical structure, $H(r)$, to be supplied. We assume $H(r)=hr$ with $h$ constant unless otherwise stated, and therefore the new input parameter is the aspect ratio, $h$. Its value is not known a priori, so we run fits over a grid of values as described below. Ultimately, we aim to constrain $h$  iteratively. Note that a constant $h$ is equivalent to assuming that the dispersion of orbital inclinations is constant across of semi-major axis. This would be the case, for example, if the disc is vertically stirred by a slightly misaligned companion and the age of the system is longer than the secular timescale \citep[e.g.][]{Wyatt1999}. It is possible that $h$ varies with radius, e.g. if the disc is self-stirred \citep[e.g.][]{Krivov2018stirring}, and we explore this possibility in \S\ref{sec:flaring}.
\end{itemize}

The remaining three parameters in \textsc{frank} are the number of brightness points, $N$, maximum radius of the fit, $R_\mathrm{max}$, and $p_0$, the scale parameter for the inverse $\Gamma$ hyperprior \citep[see][]{Jennings2020}.

\section{Testing the method on simulated data}
\label{sec:tests}

In order to test the method, we start by applying it to simulated data. This provides an opportunity to learn how the algorithm works when the true value of the aspect ratio, $h$, the inclination, $i$, and the radial structure are known. The inferences made here allow conclusions to be drawn from real data where the true radial profile is unknown.

\subsection{Simulating simple debris discs}\label{subsec:sim_simple_discs}

Simple models of discs are created by defining the distribution of the dust and the geometry of the system and then simulating images using radiative transfer simulations with the python package disc2radmc\footnote{https://github.com/SebaMarino/disc2radmc/} \citep{Marino2022} that uses RADMC3D \citep{radmc3d}\footnote{https://www.ita.uni-heidelberg.de/~dullemond/software/radmc-3d/} to produce synthetic images.

Our model consists of a Solar analog at 50~pc, surrounded by a dusty disc with a mass of 0.5~${M}_\oplus$ distributed in grains from 1~$\mu$m up to 1~cm with the same spatial distribution (see below), with a size distribution with a power law index of -3.5, and made from a mix of astrosilicates, water ice and amorphous carbon \citep[as in][]{Marino2018hd107}. To test the method, we use a range of different surface density distributions and aspect ratios, and create a model disc. For each model we create a simulated image, which is then Fourier transformed to compute its model visibilities at a set of ($u,v$) coordinates. In order to represent a realistic $uv$ coverage we use the same coverage as the ALMA observations of AU Mic presented by \cite{Daley2019}. These observations marginally resolve AU Mic's scale height, and thus we consider them as an ideal benchmark.  The visibility sampling provides a ${\sim}0\farcs3=15$~au spatial resolution at 50 pc (the distance assumed for the simulated discs) \footnote{This resolution corresponds to the beam size using natural weights of two out of three ALMA observations of AU Mic that we used here and that are reported in \citep{Daley2019}. Note that there is still information on smaller scales, with the longest baseline corresponding to 0.22}, although note that this is the image resolution and that it is slightly poorer than the true resolution power demonstrated by Frankenstein \citep{Jennings2020}. Random Gaussian noise is added to the visibilities with the same amplitude as the weights of the real AU Mic observations.

\begin{figure}
    \includegraphics[width=\linewidth]{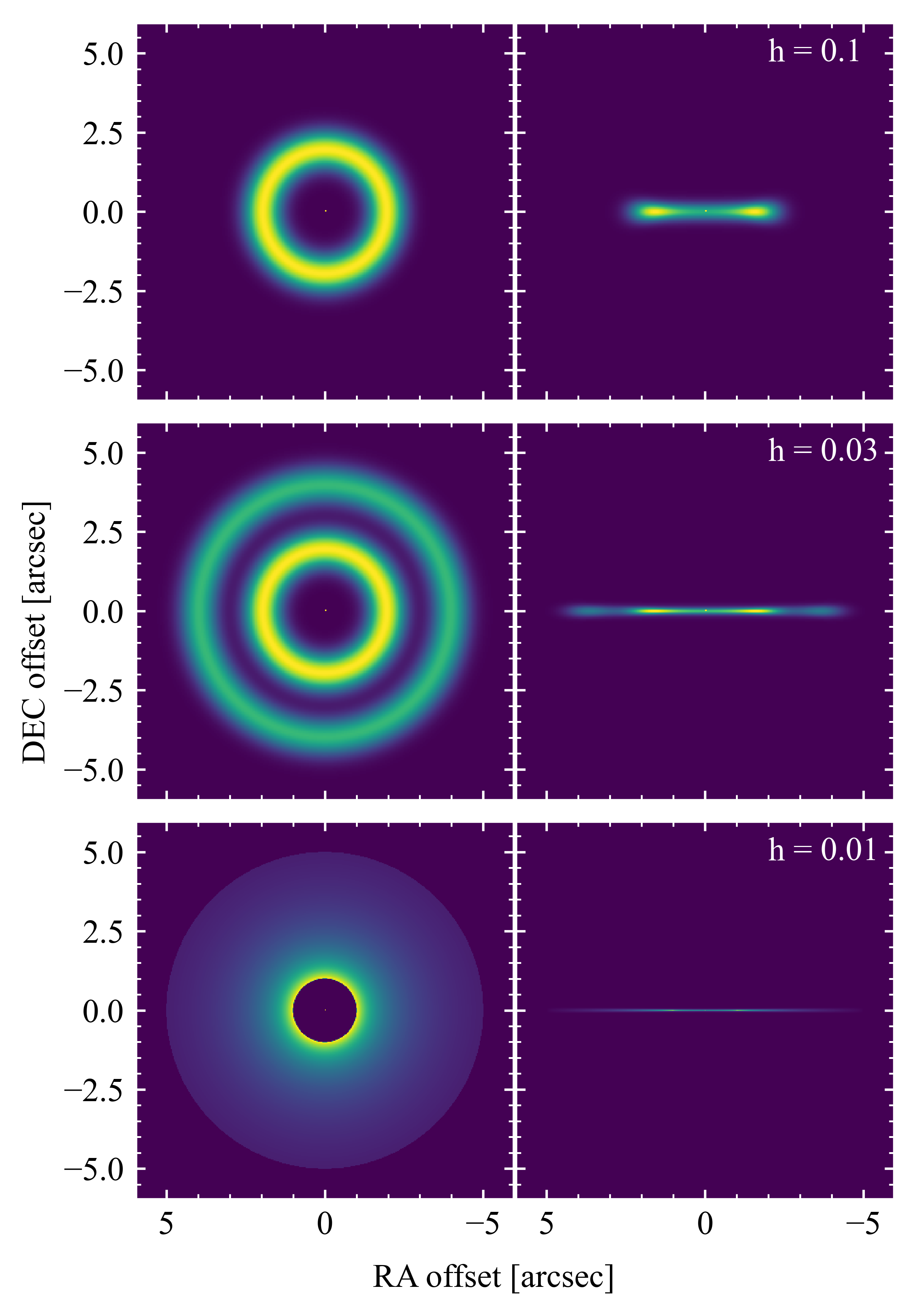}
\caption{Simulated debris disc images at 0.88 mm, assuming a Gaussian ring with $h = 0.1$, a double Gaussian ring with $ h = 0.03$, and an inverse power law distribution with $h = 0.01$ The images on the left are orientated face-on ($i = 0^{\circ}$), while the ones on the right are edge-on. These images have not been convolved with any PSF nor include noise.}
\label{fig:sim_disc_images}
\end{figure}

Three types of disc are created, each with a different type of surface density distribution: 
\begin{itemize}
    \item A set of nine discs with Gaussian radial distributions of various widths, $\sigma_r\in\{5, 10, 20\}$ au, centred at $r_c = 100$ au, and aspect ratios, $h\in\{0.01, 0.03, 0.1\}$. An example is shown in the top panel of Figure \ref{fig:sim_disc_images}, both face-on and edge-on.
    \item A case of a double Gaussian radial distribution, with $h=0.03$ and $\sigma_r = 20$ au, peaking at 100 au and 200 au, shown in the middle panel of Figure \ref{fig:sim_disc_images}.
    \item A case of a power law radial distribution ($\Sigma(r) \propto r^{-1}$) from 50 to 250~au, and with $h = 0.01$, shown in the bottom panel of Figure \ref{fig:sim_disc_images}.
\end{itemize}

\subsection{Recovering the radial profile of edge-on discs}\label{subsec:radial_profiles_of_sim}

In this section, we test how well we can recover the radial profile of our simulated edge-on disc observations with added noise using \textsc{frank}. We use the known true aspect ratio, and set $\alpha = 1.04$ and  $w_{\textrm{smooth}} = 10^{-3}$. Note that deviating from the default values ($\alpha = 1.05$ and $w_{\textrm{smooth}} = 10^{-4}$) does not change significantly the recovered profiles. To test the quality of the \textsc{frank} fits, we compare them with the true radial profiles measured using the face-on images of the discs. Note that since the central star contribution to the visibilities is simply a constant with a value equal to its flux, we subtract this from the model visibilities prior to performing the fit with \textsc{frank}. This is to avoid some oscillatory artefacts in the recovered radial profile produced as \textsc{frank} forces the model visibilities to zero instead of the stellar flux at long baselines \citep{Jennings2022}.

\begin{figure}
    \centering
    \includegraphics[width=\linewidth]{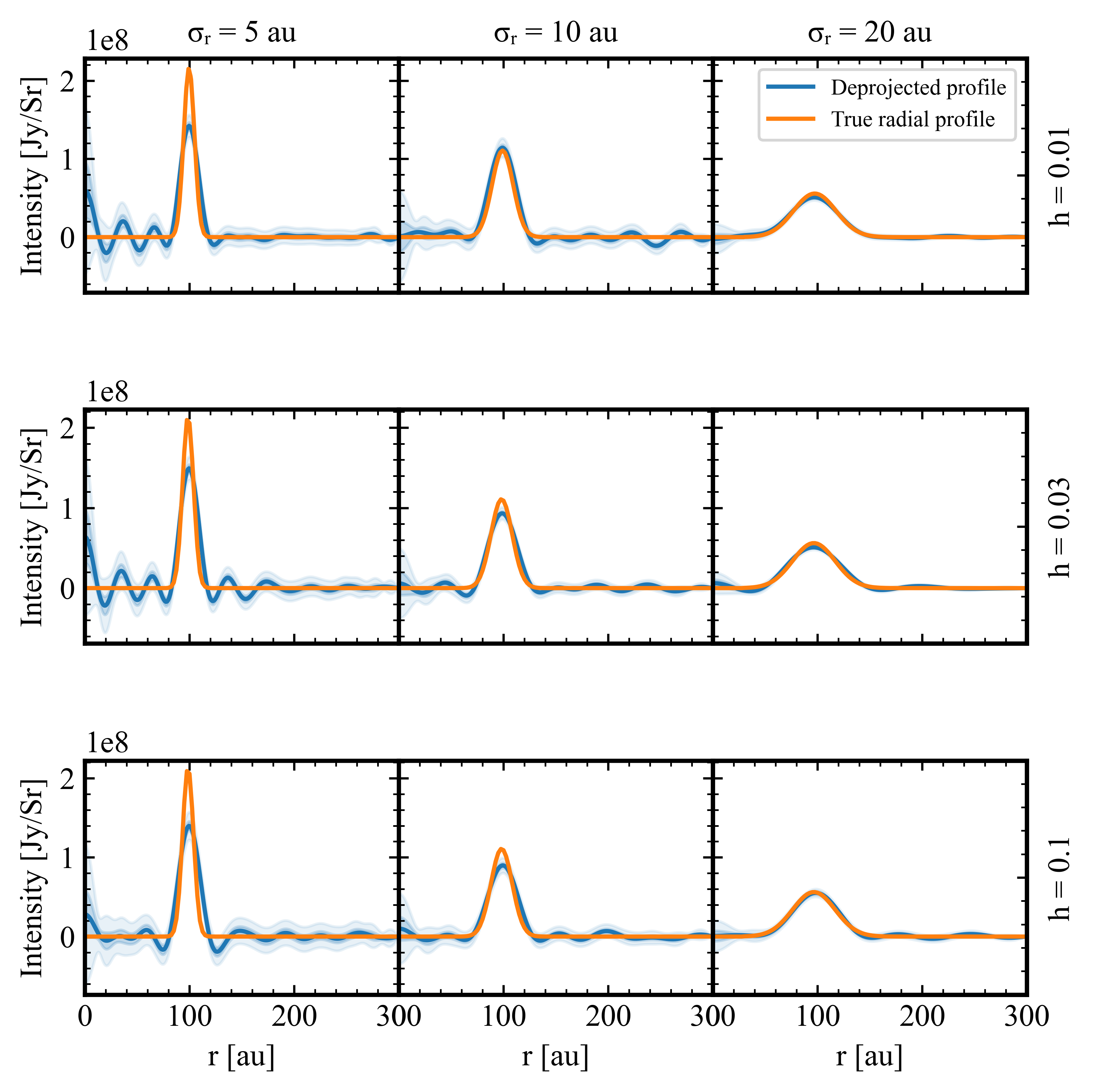}
    \caption{The recovered radial profiles of 9 simulated Gaussian debris discs, plotted in blue. The shaded regions display 1 and 3$\sigma$ confidence regions. The orange curves show the true radial profile. The aspect ratio of the discs, $h$, varies vertically; the width of the discs varies horizontally.}
    \label{fig:true_vs_debris_fits_summary}
\end{figure}

Figure \ref{fig:true_vs_debris_fits_summary} presents the radial profiles retrieved by \textsc{frank} (blue) compared with the true radial profiles (orange). The deprojected radial intensity profiles reproduce well the true profile, with a peak at $r = 100$ au, for all disc thicknesses and widths. The shape of the peak follows the true Gaussian shape closely for each fit, within 1 standard deviation (darkest shaded region) for the majority of the curve\footnote{Note that the blue shaded regions, which represents the uncertainty of the intensity profile, is computed using the diagonal of the covariance matrix. This is only an estimate that is calculated at the maximum a posteriori power spectrum that approximately represents a fit's statistical uncertainty, but it does not include the systematic uncertainty (due to incomplete $u-v$ sampling), so it is always an underestimate. For a detailed discussion see \cite{Jennings2020}}. The retrieved and true profiles only differ significantly when the ring is very narrow and sharp. In those cases, the profile recovered by \textsc{frank} has a shallower and smoother peak due to the effective resolution of Frank (set by the uv-coverage and signal-to-noise of the data). 
The deprojection remains within 3 standard deviations of zero, either side of the peak, where there is no real emission. The oscillations show a weak trend of decreasing amplitude with increasing $h$ (roughly similar for $h=0.01, 0.03$, smaller near the origin for $h=0.1$). The oscillation amplitude grows as the disc becomes increasingly narrower than the image resolution of $15$~au. This is because narrow rings in the profile generate oscillations in the corresponding visibility distribution that are not easily extrapolated by the model beyond sampled baselines. Higher resolution observations would thus reduce the amplitude of radial brightness profile oscillations.   

\begin{figure}
\centering
    \includegraphics[width=1\linewidth, trim = {0 0 0 0}, clip]{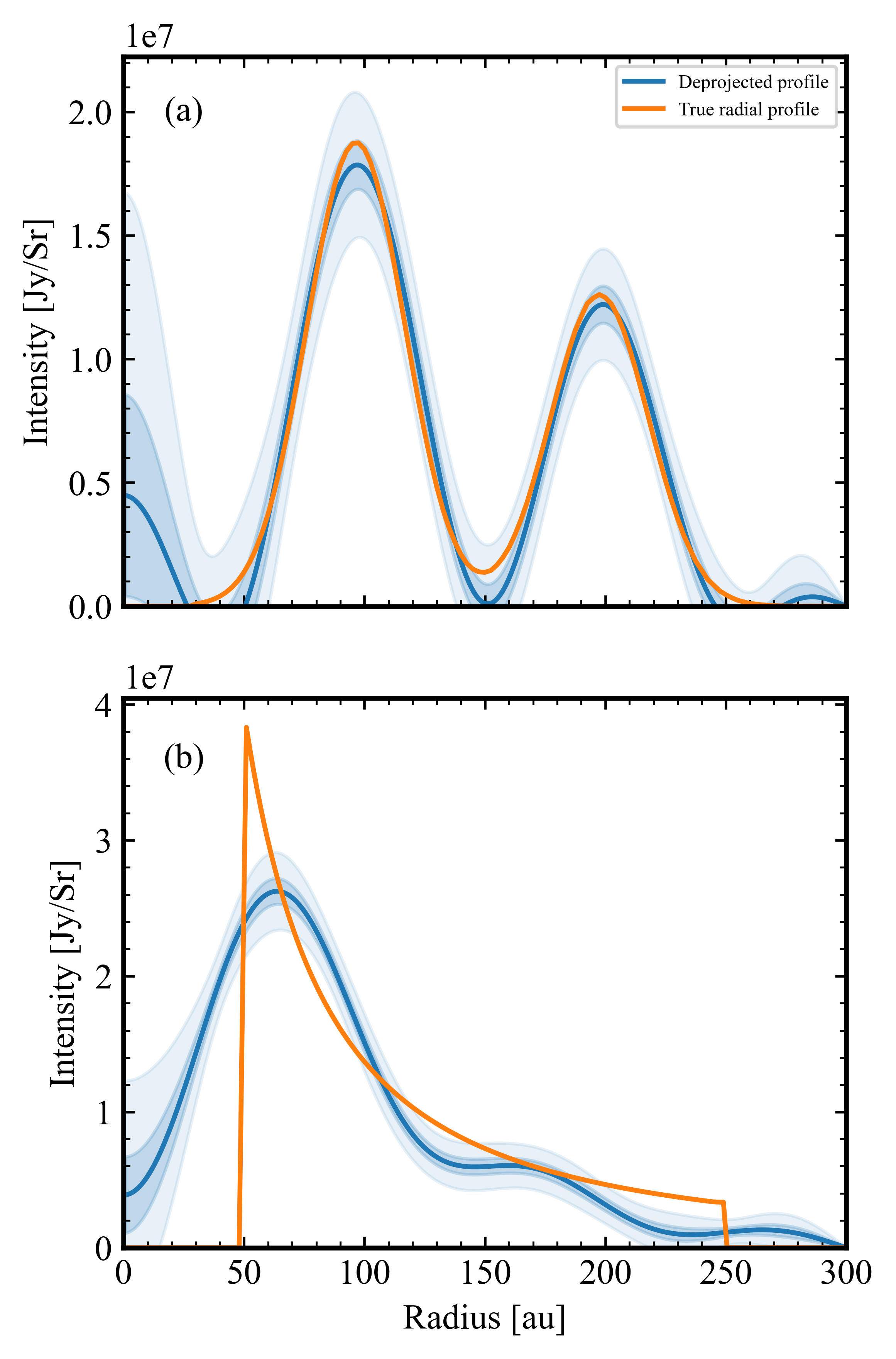}
\caption{Radial profiles recovered for radial distributions other than a single Gaussian, with the associated 1 and 3$\sigma$ confidence regions shaded, and the true radial profile shown. Panel (a) shows the profile of a simulated double Gaussian disc, where each ring has a width of 20 au. Panel (b) shows the profile of a disc simulated with a decaying distribution, $\Sigma\propto \frac{1}{r}$.} \label{fig:db_gauss_and_inv_pw_law}
\end{figure}
Figure~\ref{fig:db_gauss_and_inv_pw_law} shows \textsc{frank} can accurately recover more complex radial profiles, such as a double Gaussian distribution and an inverse power law distribution. In particular, for a double Gaussian, the fit is within 3 standard deviations of the true profile. The decaying radial profile has a sharp inner and outer edge, which \textsc{frank} smooths. The smooth decay is recovered well, although it oscillates around the true profile due to the noise and sharp inner edge that causes systematic oscillations.

To conclude, the new extension of \textsc{frank} can recover with reasonable accuracy the radial profile of simple edge-on discs whilst taking into account their vertical structure. Only sharp distributions are not well recovered, and these recovered profiles still achieve a higher resolution than standard imaging methods such as \texttt{CLEAN} \citep{Jennings2020}. 
The only difference in the radial profiles for different vertical thicknesses is that some oscillatory artefacts are stronger for small $h$. While these artifacts are pronounced in some cases, in general the algorithm performs well for all $h$.


\subsection{Estimating the aspect ratio}\label{subsec:est_h_intro}

Thus far, only the true value of $h$ has been used in order to test the ability of the algorithm to recover a radial profile of edge-on debris discs. In this section, the ability of the algorithm to constrain an unknown aspect ratio is tested. To achieve this, we need to evaluate the probability of each value of $h$, marginalized over the other parameters in the model. Ideally, we should marginalize over the brightness, $\un{I}$, the power spectrum parameters, $\un{p}$, and any geometry parameters such as the inclination and position angle.
Since $P(\un{I}|\un{V}, \un{p})$ is Gaussian in $\un{I}$, we can marginalized over $\un{I}$ analytically. This marginalization is already used in \textsc{frank} when choosing $\un{p}$, the result of which is 
\begin{align}
  \log P(\un{p} |& \un{V}, h, \boldsymbol{\beta}) = \frac{1}{2}\,\un{j}^T \unun{D}\, \un{j} + \frac{1}{2} \log|\unun{D}| - \frac{1}{2} \log |\unun{S}(\un{p})| \nonumber \\
  & - \sum_k\left[(\alpha - 1) \log p_k + \frac{p_0}{p_k} \right] - \frac{w_{\rm smooth}}{2} \log (\un{p})^T \unun{T} \log (\unun{p}) \nonumber \\
  & - \frac{1}{2}\un{V}^T\unun{N}^{-1}\un{V} + {\rm const.} \label{eqn:like_marg1}
\end{align}
\citep{Jennings2020}. Here we have explicitly denoted the dependence on the disc aspect ratio, $h$, which appears in $\unun{D}$ and $\un{j}$. We also introduce $\boldsymbol{\beta}$ to represent additional parameters such as the disc geometry and the parameters of the hyperprior,  $P(\un{p}|\alpha, w_{\rm smooth}, p_0)$, which is an inverse gamma distribution with an added smoothness term.

Next, to obtain $P(h|\un{V}, \beta)$ we need to introduce a prior on the aspect ratio (which we assume to be flat as in previous works) and to marginalize over $\un{p}$, i.e.
\begin{equation}
P(h|\un{V}, \beta) = \int P(h,\un{p}|\un{V}, \beta) {\rm d} \un{p} = \int P(\un{p}|\un{V}, h, \beta) P(h) {\rm d} \un{p}. \label{eq:P(h)}
\end{equation}
For constant $P(h)$ its value does not affect the inferred values of $h$ and has thus been dropped from subsequent expressions.

Given the complex form of \autoref{eqn:like_marg1} it is not possible to perform this marginalization analytically. Monte-Carlo methods of integrating \autoref{eqn:like_marg1} are also prohibitively expensive since $\un{p}$ contains the same number of parameters as the brightness profile, i.e. typically a few 100 parameters. To progress we therefore perform the marginalization approximately, using the Laplace method \citep{MackayBook}. That is we make a Gaussian approximation to $P(\un{p} | \un{V}, h, \boldsymbol{\beta})$ around the maximum likelihood values, $\un{p}_{\rm MAP}$. As in \citet{Jennings2020}, we maximize $P(\un{p} | \un{V}, h, \boldsymbol{\beta})$ with respect to $\log \un{p}$, and estimate the covariance from the Hessian of $\log P(\un{p} | \un{V},h, \boldsymbol{\beta})$, i.e.
\begin{align}
  \frac{{\rm d}^2 \log P(\un{p} | \un{V}, h,  \boldsymbol{\beta})}{{\rm d} \log p_k \, {\rm d} \log p_{k'}} = & \frac{1}{p_k p_{k'}}\left[\unun{Y}_{\rm f} (\mean \mean^T + \frac{1}{2} \unun{D}) \unun{Y}_{\rm f}^T\right]_{kk'} \left(\unun{Y}_{\rm f}\unun{D} \unun{Y}^T_{\rm f}\right)_{kk'} \nonumber \\
   & - \left\{ \frac{p_0}{p_k} + \frac{1}{2 p_k}\left[\unun{Y}_{\rm f} (\mean \mean^T + \unun{D}) \unun{Y}_{\rm f}^T\right]_{kk}\right\}\delta_{kk'}\nonumber \\ 
   &- w_{\rm smooth} \unun{T}_{kk'}
\end{align}
Here $\unun{Y}_{\rm f}$ is the matrix defining the Discrete Hankel Transform, as given in \citet{Jennings2020}.

We justify our use of the Laplace approximation to $P(h|\beta)$ later by comparing the values of $h$ inferred using the Laplace approximation and other metrics, including the $\chi^2$ of the best fit model and MCMC analysis of fully parametric models, finding good agreement. See \autoref{subsec:compare_to_old_h} and the Appendix for more details.

With the above definitions, we can estimate the true value of $h$, $\langle h\rangle$, as the the value that maximizes $P(h|\un{V}, \beta)$. Assuming the posterior probability behaves like a Gaussian near the maximum, we can estimate its uncertainty, $\sigma_{h}$, by solving $\log P(\langle h\rangle|\un{V}, \beta)-\log P(\langle h\rangle\pm\sigma_h|\un{V}, \beta)=1/2$. 

\subsubsection{Estimates of $h_t$ for Simulated discs}\label{subsubsec:h_est_results_for_sims}

\begin{figure}
    \includegraphics[width=\linewidth, trim = {0 0 0 0}, clip]{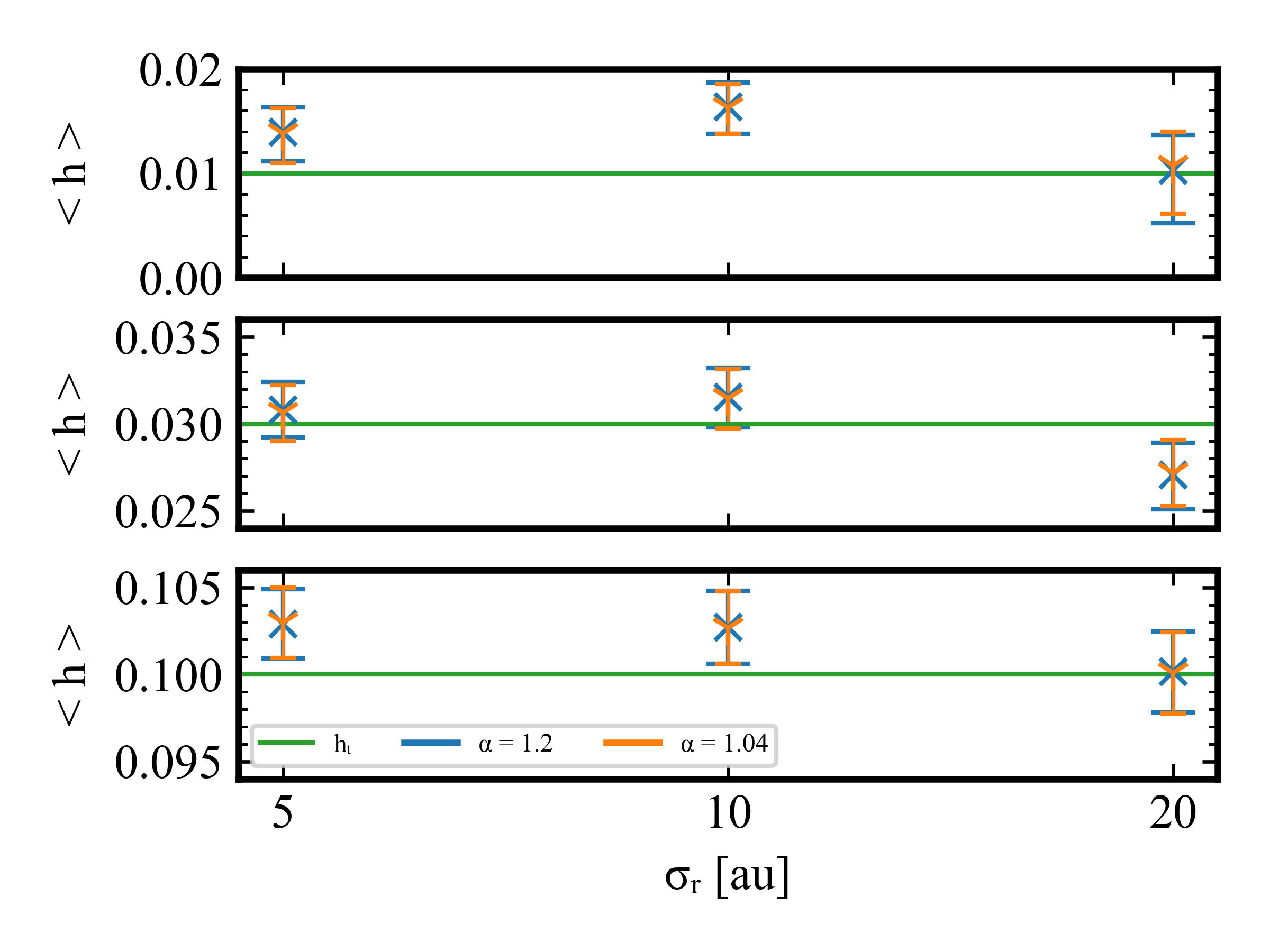}
    \caption{Aspect ratio estimates for single Gaussian discs as a function of $\sigma_r$ for three different values of $h_t$ (0.01, 0.03, and 0.1 from top to bottom). The blue and orange errorbars represent the estimates $\langle h \rangle\pm \sigma_h$ for $\alpha = 1.04, 1.2$ respectively. The green horizontal line represents $h_t$.}
    \label{fig:chisq_result_summary}
\end{figure}

In order to test how well we can retrieve $h_t$, we use the aforementioned set of nine Gaussian discs, the double Gaussian disc and power law disc. Varying the model's prior values, we find that changing $w_{\textrm{smooth}}$ from $10^{-3}$ to $10^{-4}$ produces no change in $\langle h \rangle$. We therefore fix $w_{\textrm{smooth}}$ to $10^{-3}$. We next vary $\alpha$ from $1.04$ to $1.2$, with results shown in Figure~\ref{fig:chisq_result_summary}.  




Figure~\ref{fig:chisq_result_summary} shows that all $\langle h \rangle$ estimates are close to the true value, with four estimates within $1\sigma_h$ of $h_t$, consistent with the expected 68\% for a sample this small. There is no clear trend in aspect ratio estimation accuracy with respect to the radial structure. Comparing the results of fits using $\alpha = 1.04$ and $1.2$, we find only a marginal effect on $\langle h \rangle$ and $\sigma_h$. The $\langle h \rangle$ estimate improves for lower $\alpha$, moving closer to the true value; although this effect is not shown for the widest discs, $\sigma_r = 20$ au, it becomes more apparent as the disc narrows. The size of $\sigma_h$ does not change noticeably with $\alpha$. However, for the thinnest discs ($h=0.01$), the uncertainty becomes comparable to the size of $h$ and thus $\log P$ does no longer approximate a parabola near the maximum: it decreases from the maximum slower for $h<\langle h\rangle$ than for $h > \langle h \rangle$. Therefore, an asymmetric error is found and can be seen most notably for the disc of width $\sigma_r = 20$ au.


Collectively these tests demonstrate that varying $h$ to maximize $\log P(h|\un{V}, \beta)$ -- provided the other disc properties are correct -- can produce reliable estimates on the aspect ratio of the disc. The estimate is not affected by the radial structure of the disc, although only relatively simple discs are tested. The estimated error associated with $\langle h \rangle$ reasonably encapsulates the deviation from the true value. $\langle h \rangle$ is within 1 $\sigma_h$ of $h_t$ for the majority of the performed tests and for 4/9 of the estimates shown in Figure~\ref{fig:chisq_result_summary}, illustrating that $\sigma_h$ is an appropriate choice of uncertainty. In \S\ref{subsec:compare_to_old_h} we compare our results on real data to parametric fits finding a good agreement in the estimates and derived uncertainties. Finally, we find that increasing the variance of the noise injected into our mock observations by a factor of 10 causes $\sigma_h$ to scale accordingly, by approximately $\sqrt{10}$. Therefore, $\sigma_h$ increases linearly with the noise in the data.

These and additional tests we performed also show that $\alpha$ does not have a significant effect on $\langle h \rangle$ for the simulated discs, although can worsen if $\alpha$ is too high and above the recommended range of values. We note that the discs are simple and only limited testing is performed, therefore there might be cases in which the behaviour of $\langle h \rangle$ is different. For some real debris discs, the choice of $\alpha$ does impact the ability to detect $h_t$, see \S\ref{sec:fitdata}.

\subsubsection{Threshold detection of aspect ratio}

Given the parameters of the simulated discs and uv-coverage that we assume, we conclude that we can retrieve scale heights with \textsc{frank} as small as $0.02 \arcsec$ (h=0.01 at 100 au for a system at 50 pc) using observations with a \texttt{CLEAN} beam of $0.3 \arcsec$. A detection level this small is surprising since the vertical FWHM of such a disc would be $0.05 \arcsec$, and thus only a fraction of the \texttt{CLEAN} beam. The ability of the algorithm to retrieve a very small $h$ is due to a) \textsc{frank}'s ability to achieve higher resolution than that in a \texttt{CLEAN} image, and b) the disc height only needs to be marginally resolved for the algorithm to find an estimate of $h_t$, c) our knowledge of the functional form of the vertical distribution (Gaussian and constant $h$). The algorithm has a limit at which it fails to differentiate between a very thin vertical structure and no vertical structure. Below this limit, $P(h|\un{V}, \beta)$ should not change.

\begin{figure}
\centering
    \includegraphics[width=\linewidth, trim = {0 0 0 0}, clip]{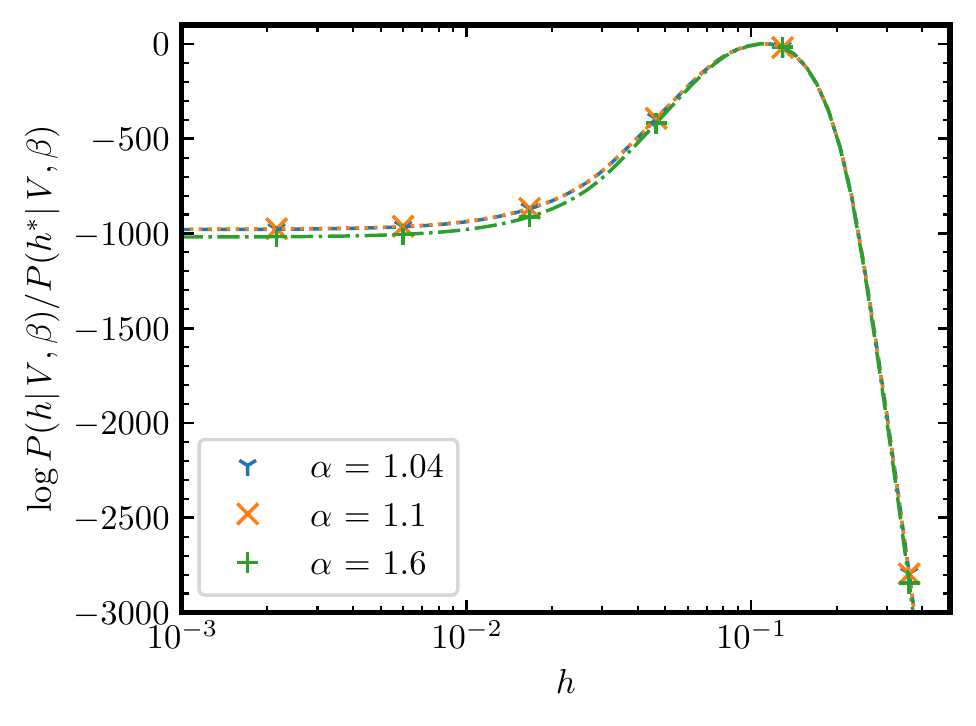}
\caption{Posterior probability distribution of $h$,  $P(h|\un{V}, \beta)$, for $h_t=0.1, \sigma_r = 5$~au and relative to its maximum $P(h^{*}|\un{V}, \beta)$. The probability distribution plateaus for a sufficiently thin disc in which the model cannot detect changes in the vertical structure; $\log\ P$ changes beyond $\approx h = 10^{-2}$, signifying that the model can detect changes in the vertical structure. The dataset used has a resolution of ${\sim}0.3 \arcsec$ \citep{Daley2019}. The discs are simulated at a distance of 50 pc and have a characteristic radius of 100 au, corresponding to $2 \arcsec$. The results show the algorithm can detect $h$  even slightly smaller than the resolution divided by the radius. $\alpha$ does not change the detection threshold level.}\label{fig:alpha_plateau}
\end{figure}

To demonstrate this threshold we estimate $h$ for a subset of the simple Gaussian discs, calculating the $P(h|\un{V}, \beta)$ for $h\in[10^{-4}, 1]$, $\alpha \in \{1.04, 1.1, 1.6\}$. Figure \ref{fig:alpha_plateau} shows $\log P(h|\un{V}, \beta)$ (relative to its maximum) for this range of aspect ratios and $\alpha$ values and $\sigma_r=5$~au. We find a clear plateau for $h$ below 0.01 for all values of $\alpha$ tested. This demonstrates the existence of a threshold for $h$ (for a given system and data set), below which $P(h|\un{V}, \beta)$ does not change.

Note that previously this interval was centred on $h_t$, as it is known. For real discs, no such centering will be possible. The plots of $\log P(h|\un{V}, \beta)$ show that if a broad initial range is used, a rough location of $h_t$ can be estimated. A narrower range, with a higher resolution, can then be used centred on the initial estimate to obtain a more precise constraint on $h$.


\subsubsection{Dependency of aspect ratio estimate on inclination}\label{subsubsec:wrong_inc}

If the inclination of the disc is not precisely known, the true vertical thickness could be obscured by inclination: assuming an inclination higher than the true value (i.e. closer to edge-on) would lead to an overestimate of vertical thickness such that the model matches the observed disc width along the minor axis. Here we want to test how well we can recover $h$ and $i$ when both are unknown. We start by considering the case that $h_t$ is known and the true inclination, $i_t$ ($90^{\circ}$ for all simulated discs), is unknown. Similar to \S\ref{subsec:est_h_intro}, we can define $ P(i|\un{V}, \beta)$ (where $\beta$ now  includes $h$) and find $\langle i \rangle$ and $\sigma_i$.
Table \ref{tab:inc_estimation_results} presents the results of estimating inclination for our nine mock Gaussian discs. For each disc the inclination estimate $\langle i\rangle$ is accurate to within $2^{\circ}$, and in all cases it is within $\sigma_i$ of $90^{\circ}$.

\begin{table}
\centering
\caption{Estimate of $i$ ($i_t=90^{\circ}$) for nine mock Gaussian discs with different vertical aspect ratios $h$ and radial standard deviations $\sigma_r$.} 
\label{tab:inc_estimation_results} 
\begin{tabular}{c|c|c}
\hline
$h$ & $\sigma_r$ (au) &  $\langle i \rangle$ ($^{\circ}$)\\
\hline
 0.01 &  5  & $89.3_{-0.5}^{+1.0}$ \\
 0.01 & 10  & $89.2_{-0.2}^{+1.2}$ \\
 0.01 & 20  & $89.6_{-0.4}^{+1.2}$ \\
0.03 &  5   & $89.6_{-0.5}^{+1.3}$ \\
0.03 & 10   & $89.5_{-0.4}^{+1.4}$ \\
0.03 & 20   & $90.0_{-0.5}^{+0.5}$ \\
0.1 &  5    & $88.5_{-1.3}^{+1.6}$ \\
0.1 & 10    & $88.5_{-1.2}^{+1.8}$ \\
0.1 & 20    & $88.9_{-1.0}^{+1.1}$ \\
\hline
\end{tabular}
\end{table}




The case above is unrealistic since if the inclination is not known precisely, it is unlikely the aspect ratio would be known beforehand. In this case where neither is known it is possible to maximize the posterior probability over both $h$ and $i$ simultaneously to obtain estimates, i.e. sample $P(h, i |\un{V}, \beta)$ in 2D over $h$ and $i$. We do this in \S\ref{subsec:hd110058} for the disc HD~110058. Here we do a simpler test where we investigate the effect of inputting an incorrect inclination on the recovered aspect ratio using one of the mock Gaussian discs. The estimate for $h_t$ is found by the same $P(h|\un{V}, \beta)$ maximisation as in \S\ref{subsubsec:h_est_results_for_sims}, however the input inclination of $90^{\circ}$ is now varied; $\langle h \rangle$ is found for $i = 80^{\circ}, 85^{\circ}, 90^{\circ}$. Note that the incorrect values of inclination break the assumption needed to claim that the maximum should be close to $h_t$ because the model is no longer correct.

\begin{figure}
\centering
    \includegraphics[width=\linewidth, trim = {0 0 0 0}, clip]{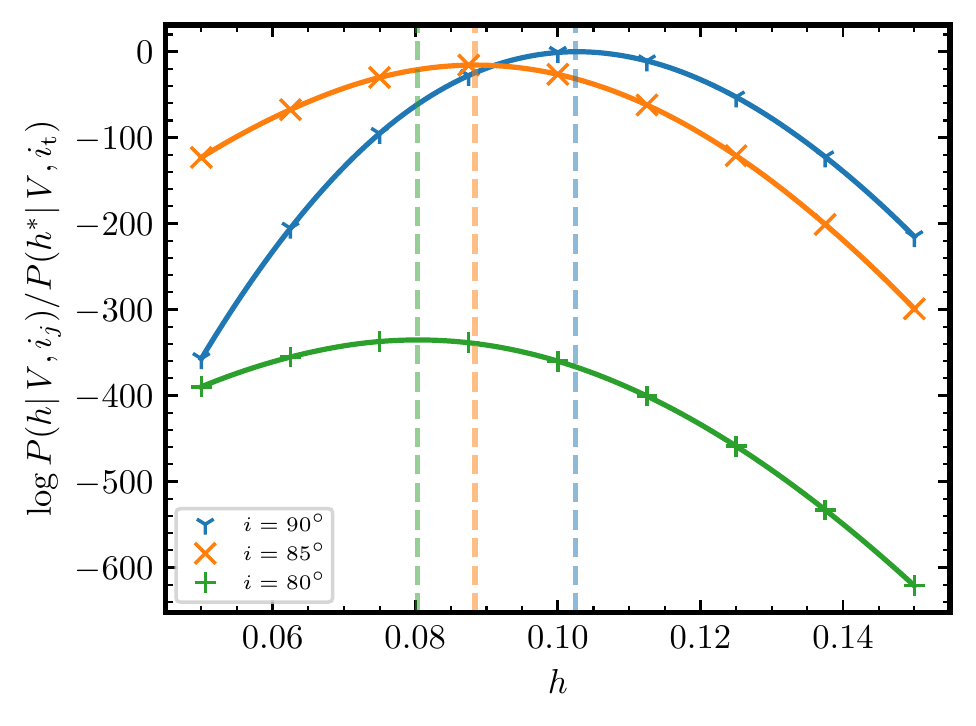}
\caption{The log of the posterior distribution as a function of aspect ratio and relative to the global maximum probability found at $h*$, for a simulated Gaussian disc with $h_t=0.1$, $\sigma_r=10$ au. The three curves correspond to different inclination assumptions: $i\in \{80^{\circ}, 85^{\circ}, 90^{\circ}\}$. The vertical dashed lines show the location of the estimate $\langle h \rangle$ found for each inclination.}
\label{fig:wrong_inc_figs}
\end{figure}

Figure \ref{fig:wrong_inc_figs} shows $\log P(h|\un{V}, \beta)$ as a function of $h$ for 3 inclinations, relative to the global maximum of the 3 curves ($P(h^{*}|\un{V}, i_{\rm t})$), for a Gaussian disc with $h_t=0.1$ and $\sigma_r = 10$ au. As the assumed inclination decreases, $\langle h \rangle$ decreases. Additionally, when the wrong inclination is provided, the $\log P(h|\un{V}, \beta)$ value at its maximum decreases significantly. Therefore, a significant decrease in $\log P(h|\un{V}, \beta)$ (here it is $\sim400$ when the inclination is incorrect by $10^{\circ}$) can indicate that an incorrect geometry has been assumed.

\section{Applying the deprojection algorithm to real data}
\label{sec:fitdata}

\begin{figure*}
\centering \includegraphics[trim=0.0cm 0.0cm 0.0cm 0.0cm, clip=true,
    width=0.8\textwidth]{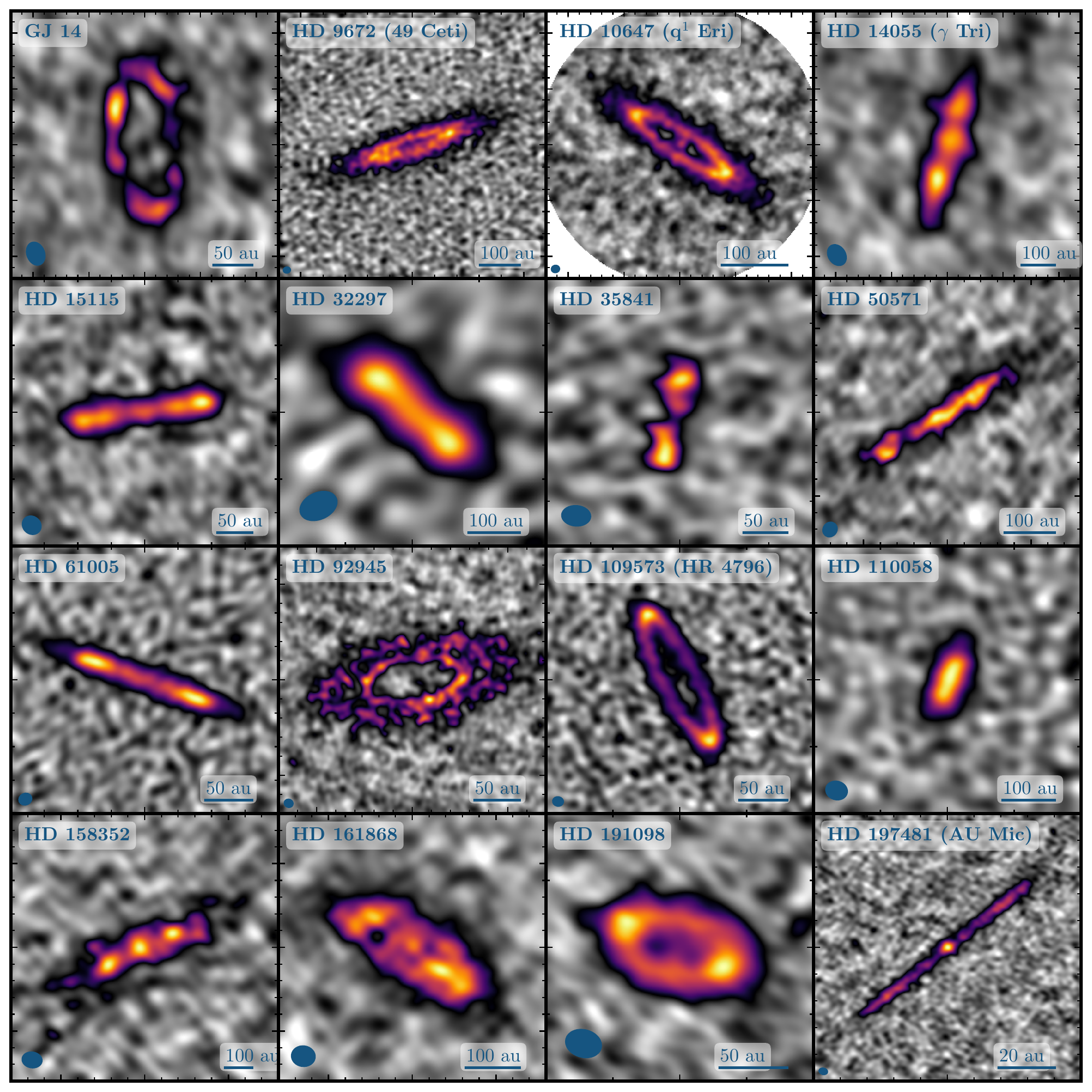}
  \caption{\texttt{CLEAN} images of the sample of studied discs. Blue ellipses at the bottom left of each image show the beam size, while the blue line at the bottom right provides a scale in au. The large and small ticks along the edges are separated by $5 \arcsec$ and $1 \arcsec$, respectively. The grey colours represent emission within $\pm3\sigma$ from zero.}
  \label{fig:almaimages}
\end{figure*}

Here we apply our method to 16 highly inclined debris discs that have been observed with ALMA with sufficient resolution to constrain the aspect ratio (at least marginally). The aim is to recover the radial intensity profile and to constrain the aspect ratio of each disc. To date, only a few debris discs have had their vertical structure constrained at (sub-)mm wavelengths. These include AU Mic  \citep{Daley2019, Vizgan2022}, $\beta$ Pic \citep{Matra2019betapic}, HD~110058 \citep{Hales2022}, HD~16743 \citep{Marshall2023}, and marginally for HD~92945  \citep{Marino2019} and HR~4796 \citep{Kennedy2018}. These measurements, however, relied on using parametric models to fit the visibilities. Recently, \cite{Han2022} constrained the vertical structure of AU Mic using a non-parametric model in the image plane.   
    
As part of the REASONS survey, Matr\'a et al. (in prep) collected and analysed most of the ALMA observations of debris discs and fit them uniformly using a parametric model consisting of Gaussian radial and vertical profiles. Based on that sample and results, we collected the ALMA dust continuum data of the 16 highly inclined debris discs for which Matr\'a et al. found a constraint on $h$ (rather than just an upper limit) by fitting a parametric model: HD~9672 \citep[49~Ceti band 8 data,][]{Higuchi2019}, HD~10647 \citep[$q^1$~Eri band 7 data,][]{Lovell2021}, HD~15115 \citep[band 6 data,][]{MacGregor2019}, HD~32297 \citep[band 6 data,][]{MacGregor2018}, HD~61005 \citep[band 6 data,][]{MacGregor2018}, HD~92945 \citep[band 7 data,][]{Marino2019}, HD~109573 \citep[HR~4796,][]{Kennedy2018}, HD~110058 \citep[band 7 data,][]{Hales2022}, HD~197481 \citep[AU~Mic band 6 data,][]{Daley2019}, and data from the REASONS survey of GJ~14, HD~14055 ($\gamma$~tri), HD~35841, HD~50571, HD~158352, HD~161868, HD~191089 \citep[][Matr\`a et al. in prep]{Sepulveda2019}. The \texttt{CLEAN} images of these sources are presented in Figure~\ref{fig:almaimages}. 



The disc inclination and position angle have been well constrained by Matr\`a et al. (in prep) for most of these sources through an MCMC approach combining parametric models, radiative transfer simulations, and Fourier transforming synthetic images to fit the ALMA visibilities. Matr\`a et al. (in prep) considered a simple Gaussian radial profile, which is sufficient for most discs analyzed here, and thus we adopt these values in our modelling except for HD~92945. This disc has a more complex radial structure with a gap \citep{Marino2019}, and thus we use the values derived by \cite{Marino2021} that used the same approach but considering a more complex radial profile. In one case, HD~110058, the disc inclination is poorly constrained by the ALMA observations alone \citep{Hales2022}. Therefore, for this source, we use the PA derived by Hales et al. and treat the disc inclination as a free parameter that also needs to be varied to constrain $h$. A 2D sampling of $P(h, i |\un{V}, \beta)$ is performed in \S\ref{subsec:hd110058} and the radial profiles are analysed. The method for estimating the aspect ratio of 15 of the 16 debris discs is the same as used for the simulated discs: vary $h$ and find the value that maximises $P(h|\un{V}, \beta)$. Note that prior to sampling $P(h|\un{V}, \beta)$, we re-scale the uncertainties of the visibilities by a factor such that the reduced $\chi^2$ is equal to 1. This is due to the uncertainty on the visibilities (or weights) having the right relative magnitudes, but typically being erroneous by a small factor close to 1.8 \citep[see][for more information]{Marino2018hd107, Matra2020, Marino2021}.
 
The aspect ratios and profiles presented are, in general, retrieved with \frank{} parameters of $\alpha = 1.04$ and $w_{\textrm{smooth}} = 10^{-3}$, and with the outer radius of the fit $R_{\textrm{out}} \approx 1.5 - 2\times$ the disc's outer edge.  We tested a range of parameters to see how the radial features vary and find that these values produce optimal fits for most of the discs in our sample. The features recovered in the radial profiles are also robust to small changes in these parameters. Two exceptions are HD~191089 and HD~110058, where for $\alpha \geq 1.04$  significant negative artefacts were produced in the radial profiles. We found more sensible fits in these cases when reducing $\alpha$ to 1.02 and 1.01, respectively, and increasing $w_{\textrm{smooth}}$ to $10^{-2}$ for HD~110058.


The final radial profiles are recovered using the estimated aspect ratio. Note that the flux of the central star, analogous to the simulated data, is subtracted from the measured visibilities before performing a fit with \frank{}. The stellar flux is obtained from the best fit value of a parametric fit by Matr\`a et al. (in prep) and \cite{Marino2021}.

\subsection{Estimates of the aspect ratio of debris discs}\label{subsec:real_h_estimation}

The estimates of $h$ for the 16 debris discs are in Table \ref{tab:real_h_results}, including the estimate for HD~110058, analysed in \S\ref{subsec:hd110058}. The detection levels are separated into three classes: the result is `significant' if the best aspect ratio estimate is more than $3\sigma$ from $h=0$; `marginal' when between $1-3\sigma$; and a `limit' estimate when within 1$\sigma$ from $h=0$, in which case we quote a 3$\sigma$ upper limit. Using \textsc{frank} we find 9 significant estimates of $h$, 5 marginal estimates, and only two limit estimates: HD~14055 and HD~191089. The estimates of $h$ range from 0.02 (AU Mic) to 0.18 (HD 158352) and 0.21 (HD~110058), with a median aspect ratio amongst significantly and marginally detected discs of 0.05. These non-limit cases have a median error of 14\%, with the tightest constraint being placed on the aspect ratio of HD~109573, at only 5\%. The two limit estimates for the aspect ratio, HD14055 and HD~191089, have upper limits of 0.1 and 0.19, respectively, that are consistent with the estimates for the bulk of our sample.


\begin{table}
\caption{Estimates of $h$ for the 16 analysed discs. The detection levels are classified as significant when the best fit is more than $3\sigma$ from $h=0$; marginal when between $1-3\sigma$; and a limit when it is within 1$\sigma$ from $h=0$, in which case we quote a 3$\sigma$ upper limit. The inclination, position angle and stellar flux values are from Matr\`a et al. (in prep), with the exception of HD~92945 \citep{Marino2021} that included a disc gap in its modelling. HD~50571's inclination is constrained to $>80^{\circ}$, and thus we assume $i=85^{\circ}$ for this system. HD~110058 is analysed in \S\ref{subsec:hd110058}, and we quote the estimate for $i>80^{\circ}$. The second column shows the distance of each target from \citet{Gaiadr3}.}
\setlength\tabcolsep{4.0pt}
\begin{tabular}{lrcrrlll}
\hline
\hline
System   & d & $i$  & PA  & $F_{\star}$  & Detection & $\alpha$ & $\langle h \rangle$         \\ 
System   & pc & [$^{\circ}$] & [$^{\circ}$] &  [$\mu$Jy] &  &  &         \\ 
\hline
AU Mic  & 9.7 & 88.4 & 128.5 &  320 & Significant & 1.04     & $0.020_{-0.002}^{+0.002}$\\
GJ 14 & 14.7 & 64.0 & 5.0 & 40 & Marginal & 1.04     & $0.05_{-0.04}^{0.02}$ \\
HD 9672 & 57.2 & 79.1 & 107.4 & 100 & Significant & 1.04 & $0.050_{-0.007}^{+0.007}$\\
HD 10647 & 17.3 & 77.2 & 56.8 & 170 & Marginal & 1.04     & $0.037_{0.008}^{0.007} $\\
HD 14055 & 35.7 & 81.1 & 163.3 & 0 & Limit & 1.04& $< 0.10$ \\
HD 15115 & 48.8 & 88.0 & 98.5 & 40 & Significant & 1.04     & $0.048_{-0.007}^{+0.007}$   \\
HD 32297 & 129.7 & 87.0 & 47.8 & 80 & Significant & 1.04     & $0.08_{-0.01}^{+0.01}$      \\
HD 35841 & 103.1 & 84.0 & 167.0  & 0 & Marginal & 1.04 & $0.15_{-0.06}^{+0.05}$ \\
HD 50571 & 33.9 & 85.0 & 121.9 & 40 & Significant & 1.04 & $0.11_{-0.02}^{+0.02}$\\
HD 61005 & 36.5 & 85.7 & 70.3 & 0 & Significant & 1.04     & $0.039_{-0.004}^{+0.003}$   \\
HD 92945 & 21.5 & 65.4 & 100.0  & 35 & Marginal & 1.04     & $0.04_{-0.01}^{+0.01}$     \\
HD 110058 & 130.1 & $>80$ & 157.0 & 5 & Significant & 1.01 & $0.21_{-0.03}^{0.03}$ \\
HD 109573 & 70.8 & 76.5 & 26.7 & 70 & Significant & 1.04 & $0.052_{-0.003}^{+0.003}$ \\
HD 158352 & 63.8 & 81.0 & 114.0 & 0 & Significant & 1.04 & $0.18_{-0.02}^{+0.02}$\\
HD 161868 & 29.7 & 68.0 & 57.0 & 50 &  Marginal & 1.04     & $0.15_{-0.04}^{+0.03}$      \\
HD 191089 & 50.1 & 60.0 & 73.0 & 40 & Limit & 1.02     & $<0.19$  \\ 
\hline
\end{tabular}\label{tab:real_h_results}
\end{table}

\subsubsection{Distribution of vertical thickness}\label{subsubsec:real_disc_summary}


\begin{figure}
    \centering
    \includegraphics[width=\linewidth]{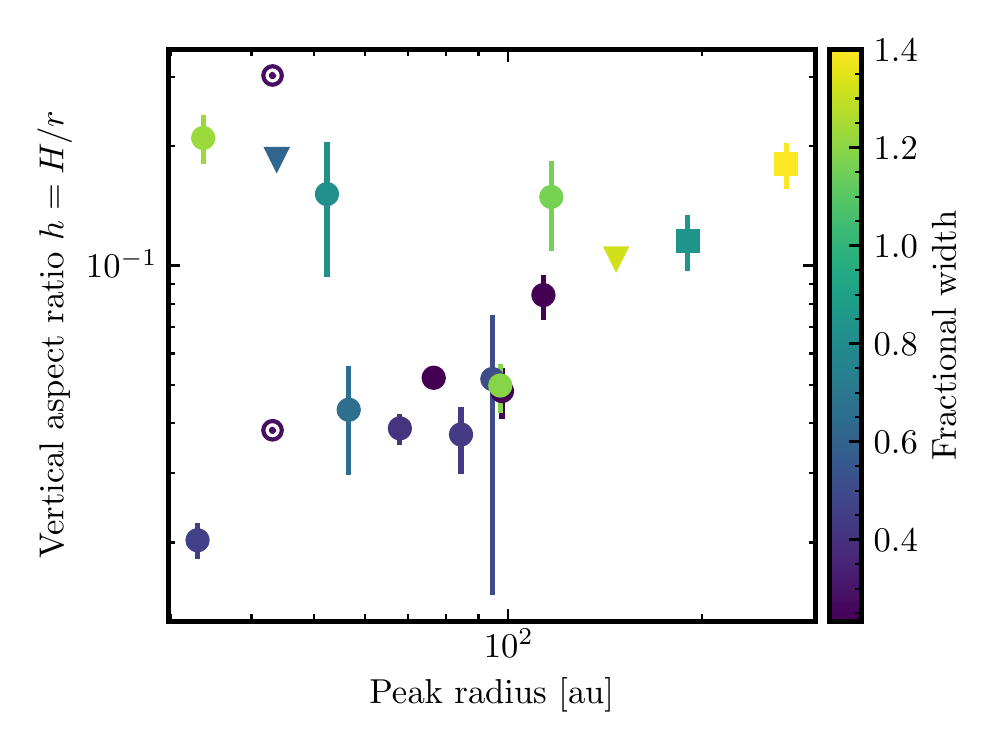}
    \includegraphics[width=\linewidth]{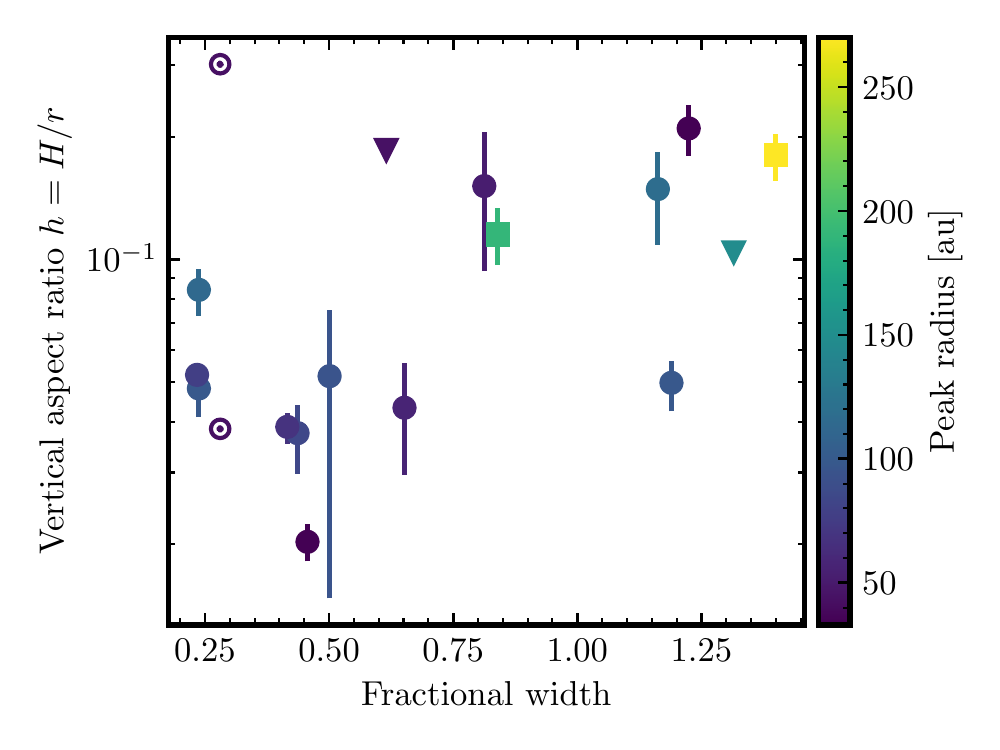}
    \caption{ The estimate of the aspect ratio of each debris disc is shown against the radius of maximum intensity (top) and fractional width (bottom). Points are colored by the disc's fractional width or radius of maximum intensity, respectively. The peak radius of HD~158352 and HD~50571 is not resolved by \textsc{frank}, so we use their peak radius and fractional width estimated by fitting a parametric model (Matrà et al. in prep). These points are marked with squares. For HD~110058, we use the fit found for $i=90^{\circ}$, $h = 0.22$. The "$\odot$" symbol represents the classical Kuiper belt with two $h$ values corresponding to its dynamically cold and hot components \citep{Brown2001}, and fractional width derived using the L7 synthetic unbiased model of the Kuiper belt \citep{Kavelaars2009, Petit2011}. }
    \label{fig:real_disc_sum}
\end{figure}

\textsc{frank} reveals a variety of radial and vertical structures in these sources. The properties of the discs are summarised in Figure \ref{fig:real_disc_sum} (along with the results in \S\ref{subsec:hd110058} for HD~110058), where we show the estimated aspect ratio as a function of the disc's peak radius (colour-coded by its fractional width, top) and as a function of the disc fractional width (colour-coded by its peak radius, bottom). The fractional width is defined as
\begin{equation}
    \textrm{Fractional width }=\frac{r_{\textrm{max}/2}^+-r_{\textrm{max}/2}^-}{r_{\textrm{max}}},\label{eqn:fwidth}
\end{equation}
where $r_{\textrm{max}}$ is the radius of maximum intensity, and $r_{\textrm{max}/2}^{\pm}$ are the radii where the intensity is half the maximum. The "$\odot$" symbol represents the classical Kuiper belt with two $h$ values corresponding to its dynamically cold and hot components \citep{Brown2001}. The classical Kuiper belt's peak radius and fractional width are derived from the L7 synthetic unbiased model of the Kuiper belt \citep{Kavelaars2009, Petit2011}.

We find no tight correlation between $h$ and the peak radius or fractional width. However, our sample has a lack of belts with a large peak radius and small $h$ and a tentative bimodal distribution of $h$, with a valley going from small fractional widths and large aspect ratios to large fractional widths and small aspect ratios. We note that our sample is biased since it only contains systems for which existing observations and parametric modeling led to a constraint in $h$ (Matrà et al. in prep). Therefore, these results must be taken with caution. If this remains a trend with a larger and less biased sample, it could indicate the presence of two separate mechanisms for vertically stirring debris discs up to different levels. For example, self-stirring or secular interactions could be responsible for values of $h$ below 0.06 \citep{Matra2019betapic}, while planet scattering could produce higher $h$ values \citep{Nesvorny2015} and at the same time widen belts leading to high fractional widths. The cold classical Kuiper belt (lower "$\odot$" symbol) fits with the low $h$ and low fractional width population. The hot classical Kuiper belt, however, has a high $h$ value but a low fractional width making it an outlier of the high $h$ population. This low fractional width might be misleading as the Kuiper belt has other hot populations (e.g. scattered and resonant populations) that have a wider radial distribution and would push the fractional width to higher values and closer to the high $h$ and high fractional width population of exoKuiper belts.




\subsection{Deprojected radial profiles of debris discs with known inclination}\label{subsec:real_profiles}

\begin{figure*}
\centering
\begin{subfigure}{.33\textwidth}
\centering
    \includegraphics[width=\linewidth, trim = {0 0 0 0}, clip]{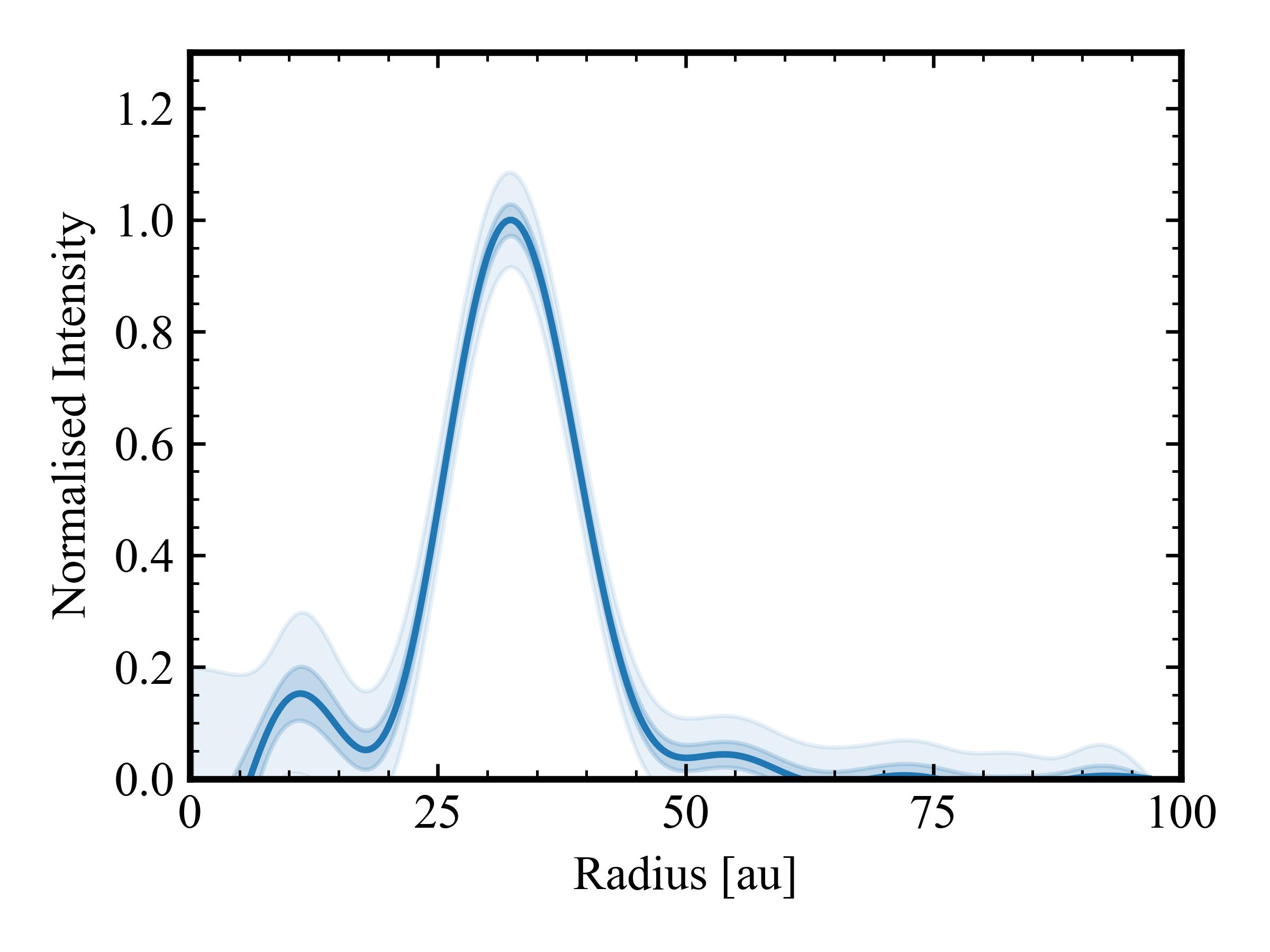}
    \caption{AU Mic}
    \label{fig:aumic}
\end{subfigure}%
\begin{subfigure}{.33\textwidth}
\centering
    \includegraphics[width=\linewidth, trim = {0 0 0 0}, clip]{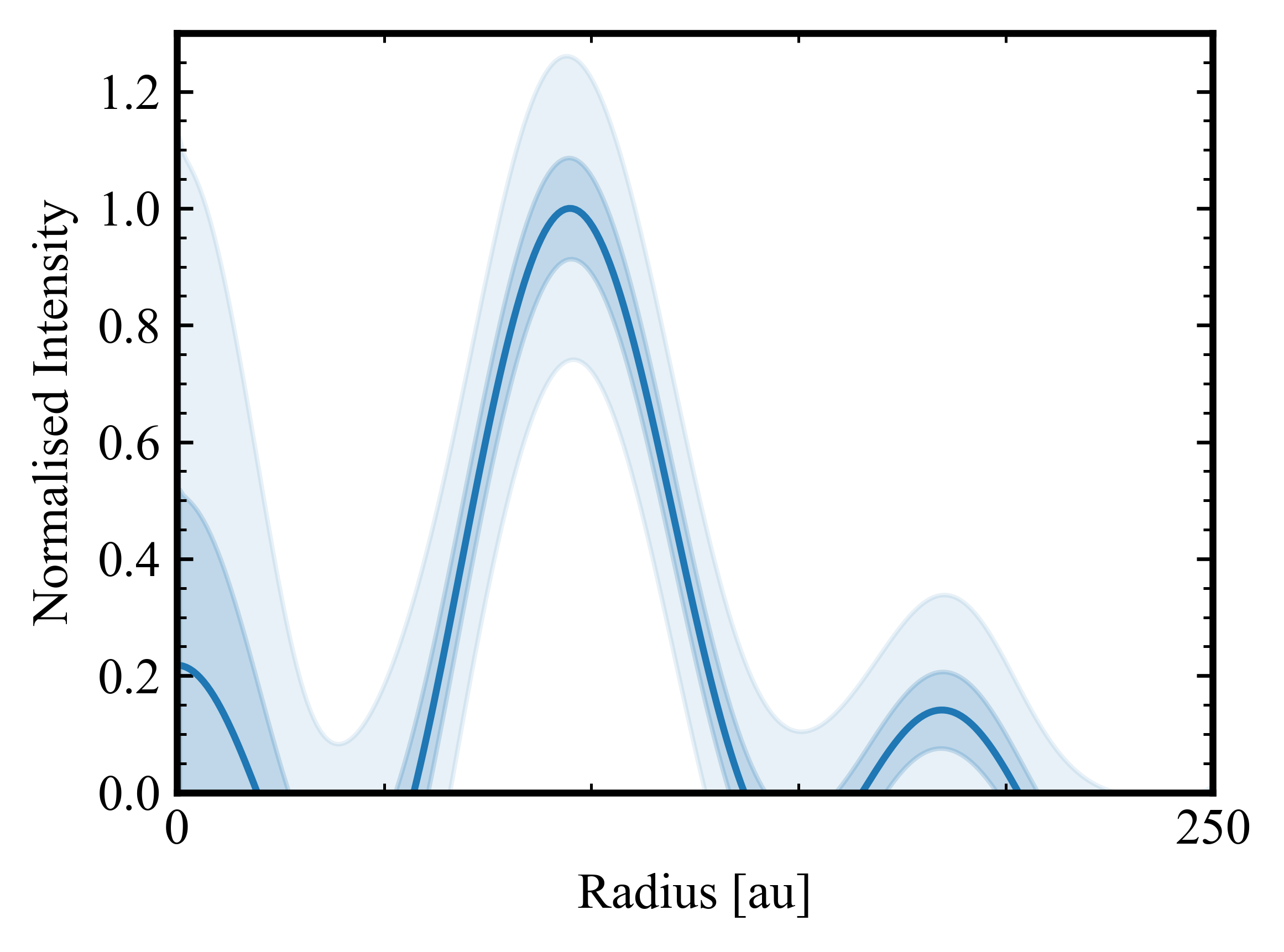}
    \caption{GJ 14}
    \label{fig:gj14}
\end{subfigure}%
\begin{subfigure}{.33\textwidth}
\centering
\includegraphics[width=\linewidth,  trim = {0 0 0 0}, clip]{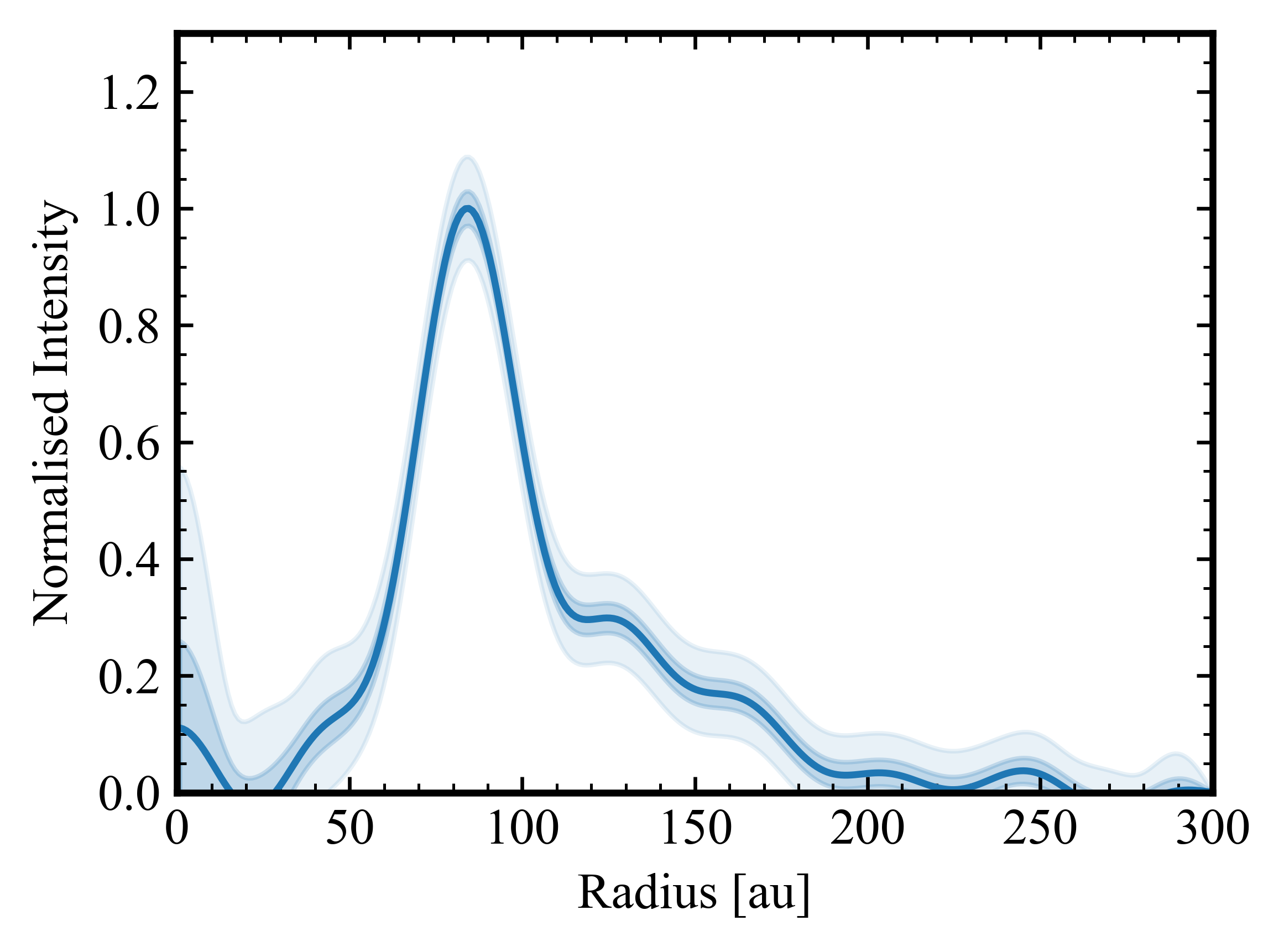}
    \caption{HD 10647}
    \label{fig:hd10647}
\end{subfigure}
\begin{subfigure}{.33\textwidth}
\centering
    \includegraphics[width=\linewidth, trim = {0 0 0 0}, clip]{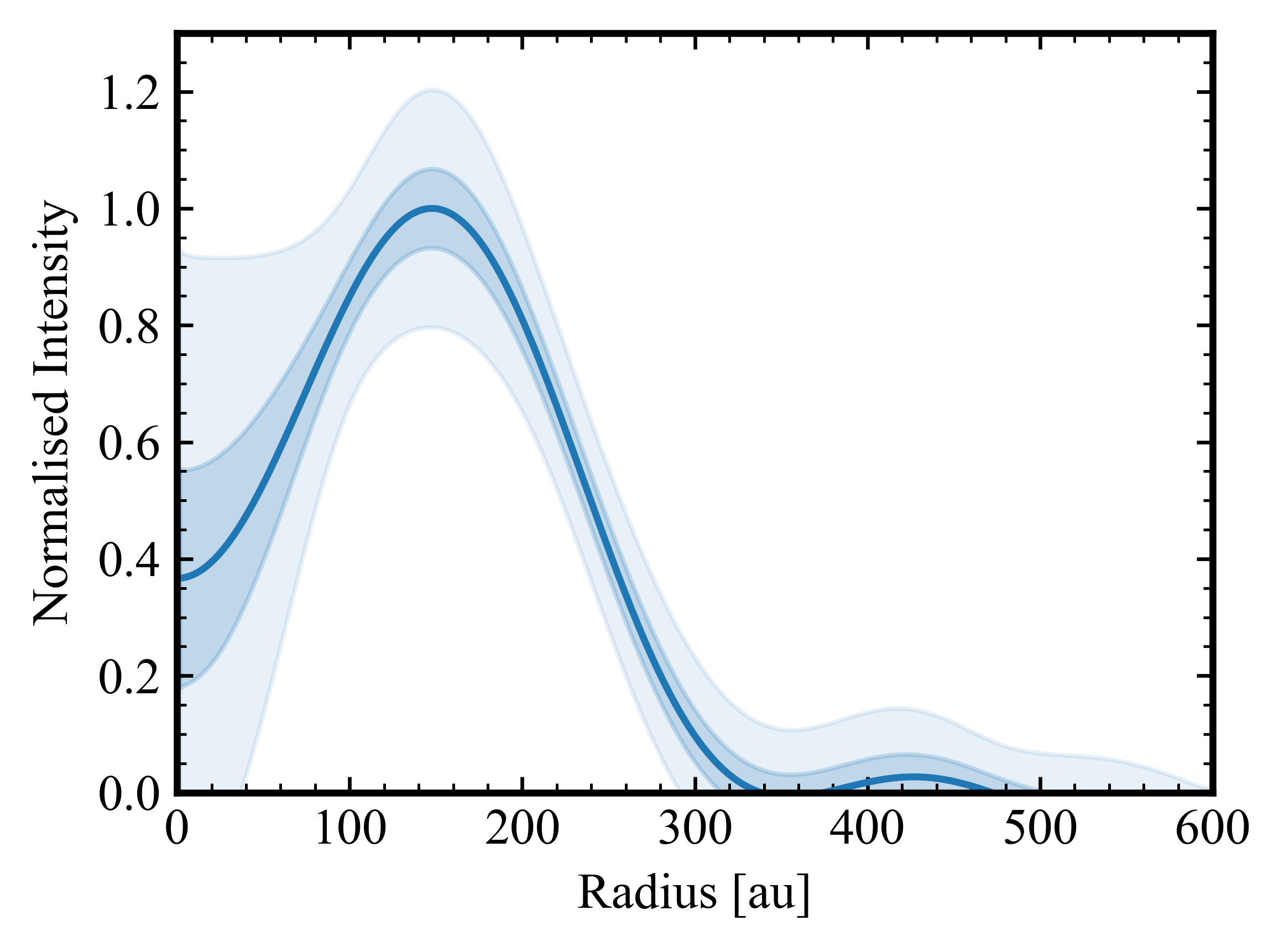}
    \caption{HD 14055}
    \label{fig:hd14055}
\end{subfigure}%
\begin{subfigure}{.33\textwidth}
\centering
    \includegraphics[width=\linewidth, trim = {0 0 0 0}, clip]{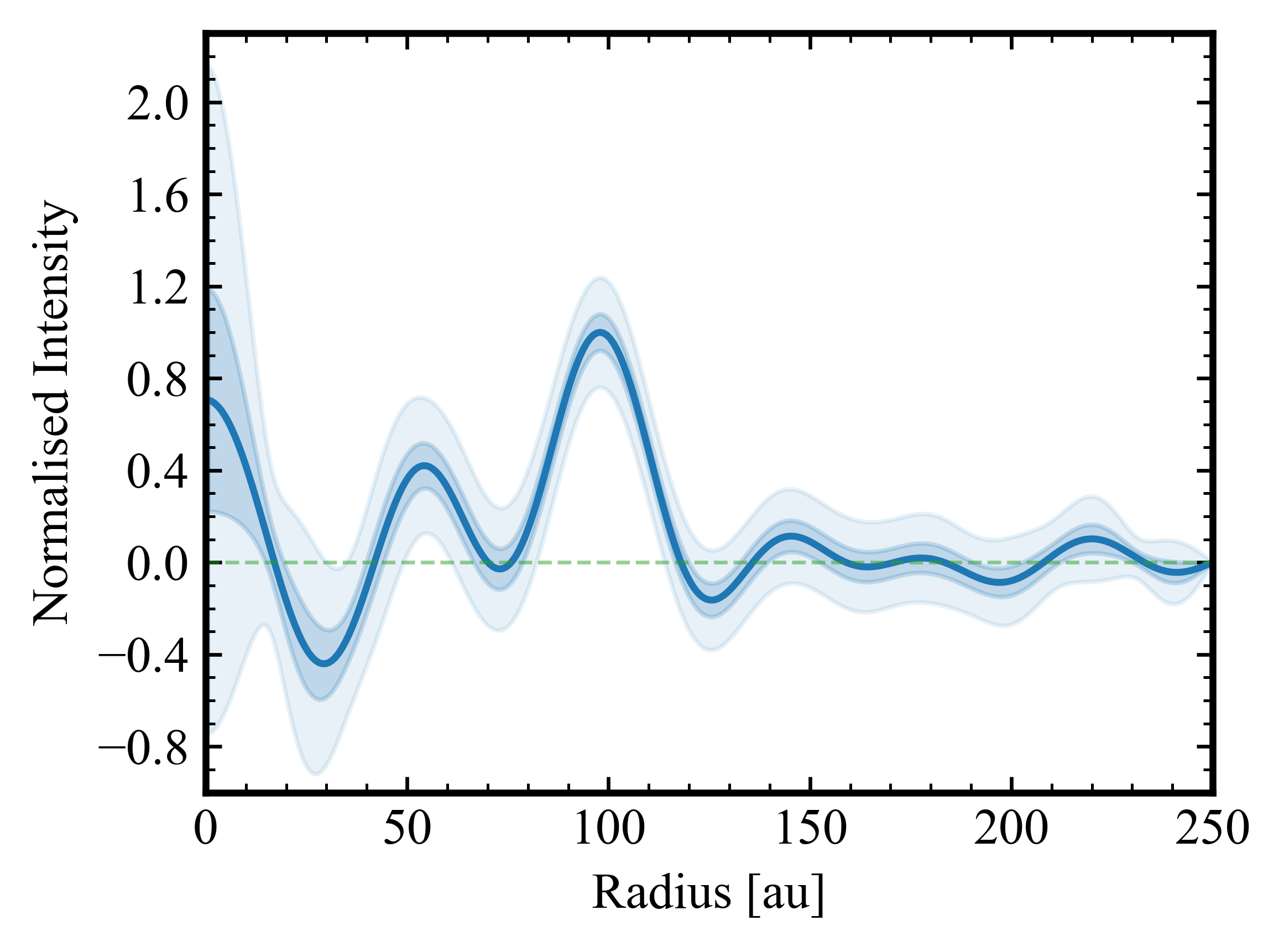}
    \caption{HD 15115}
    \label{fig:hd15115}
\end{subfigure}%
\begin{subfigure}{.33\textwidth}
\centering
    \includegraphics[width=\linewidth, trim = {0 0 0 0}, clip]{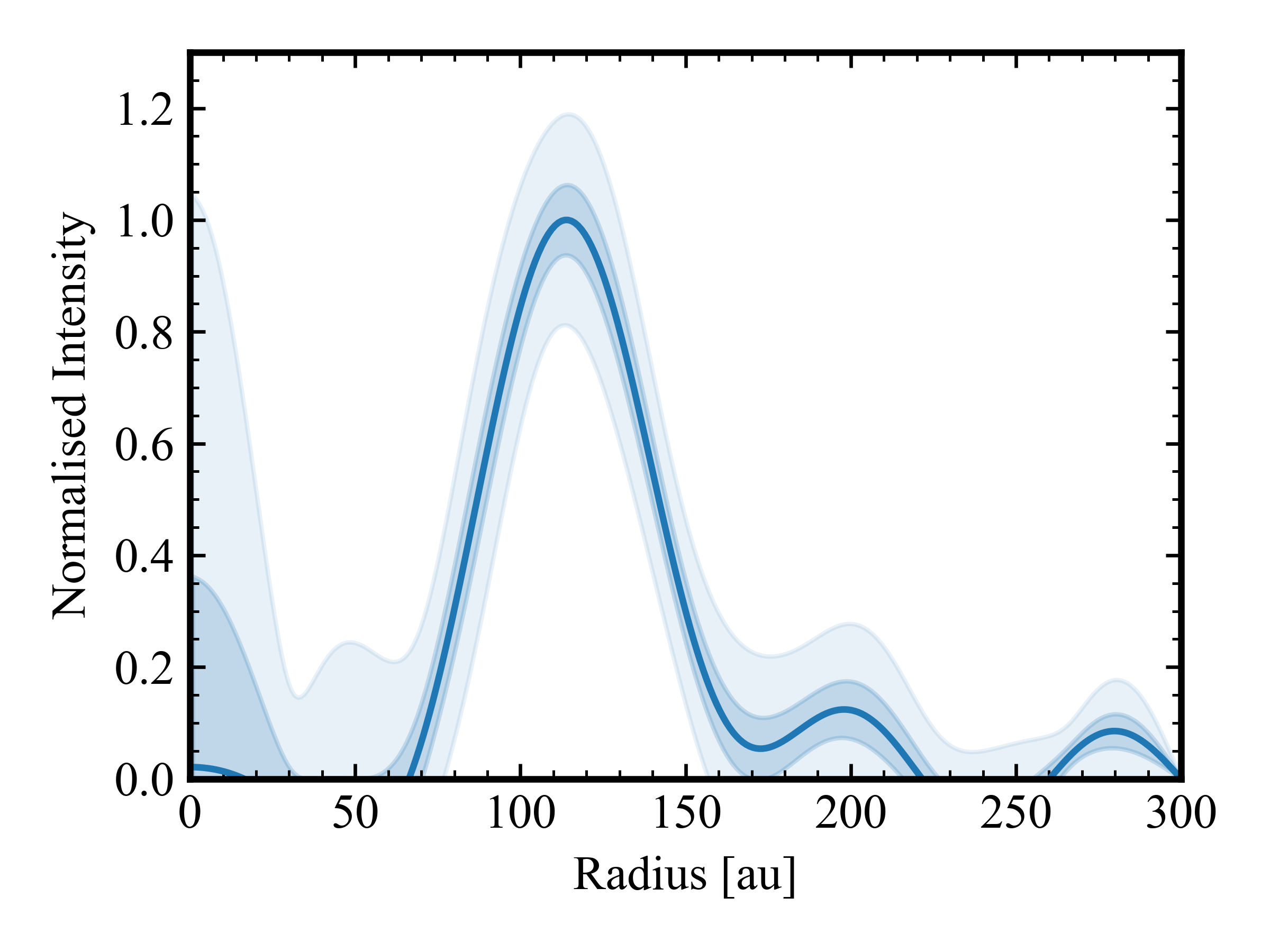}
    \caption{HD 32297}
    \label{fig:sub1501}
\end{subfigure}
\begin{subfigure}{.33\textwidth}
\centering
\includegraphics[width=\linewidth,  trim = {0 0 0 0}, clip]{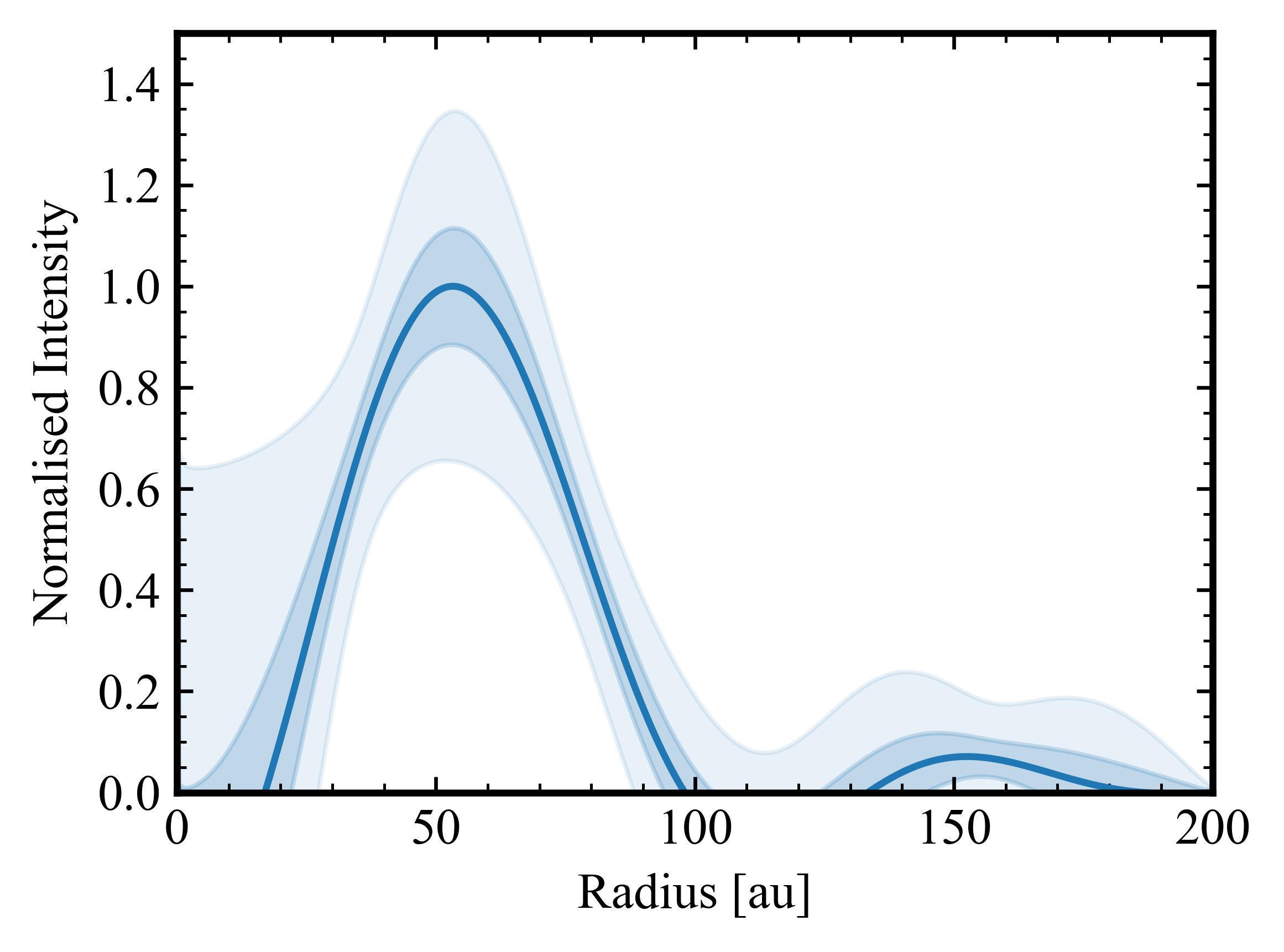}
    \caption{HD 35841}
    \label{fig:hd35841}
\end{subfigure}%
\begin{subfigure}{.33\textwidth}
\centering
    \includegraphics[width=\linewidth, trim = {0 0 0 0}, clip]{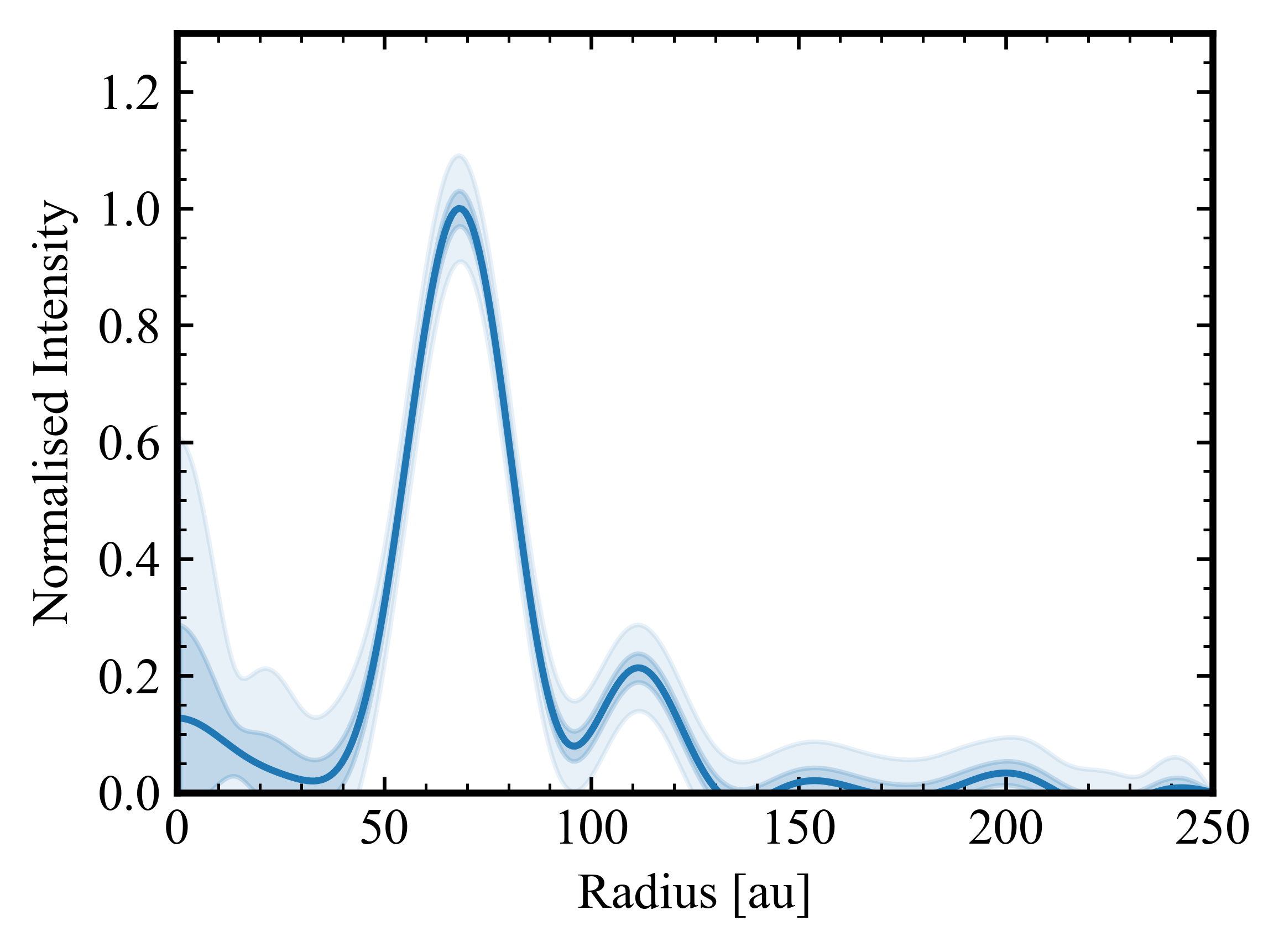}
    \caption{HD 61005}
    \label{fig:hd61005}
\end{subfigure}%
\begin{subfigure}{.33\textwidth}
\centering
    \includegraphics[width=\linewidth, trim = {0 0 0 0}, clip]{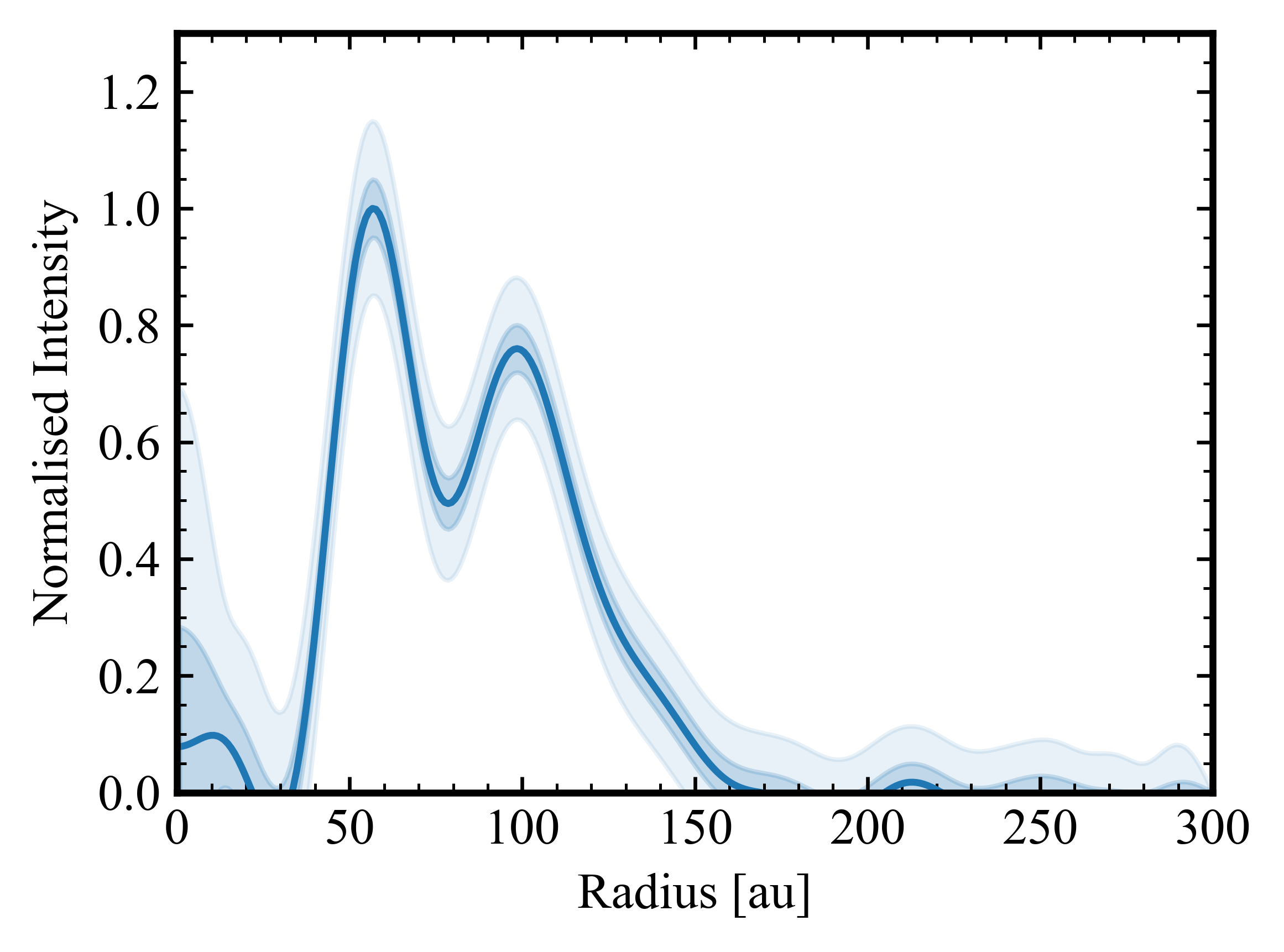}
    \caption{HD 92945}
    \label{fig:hd92945}
\end{subfigure}
\begin{subfigure}{.33\textwidth}
\centering
\includegraphics[width=\linewidth,  trim = {0 0 0 0}, clip]{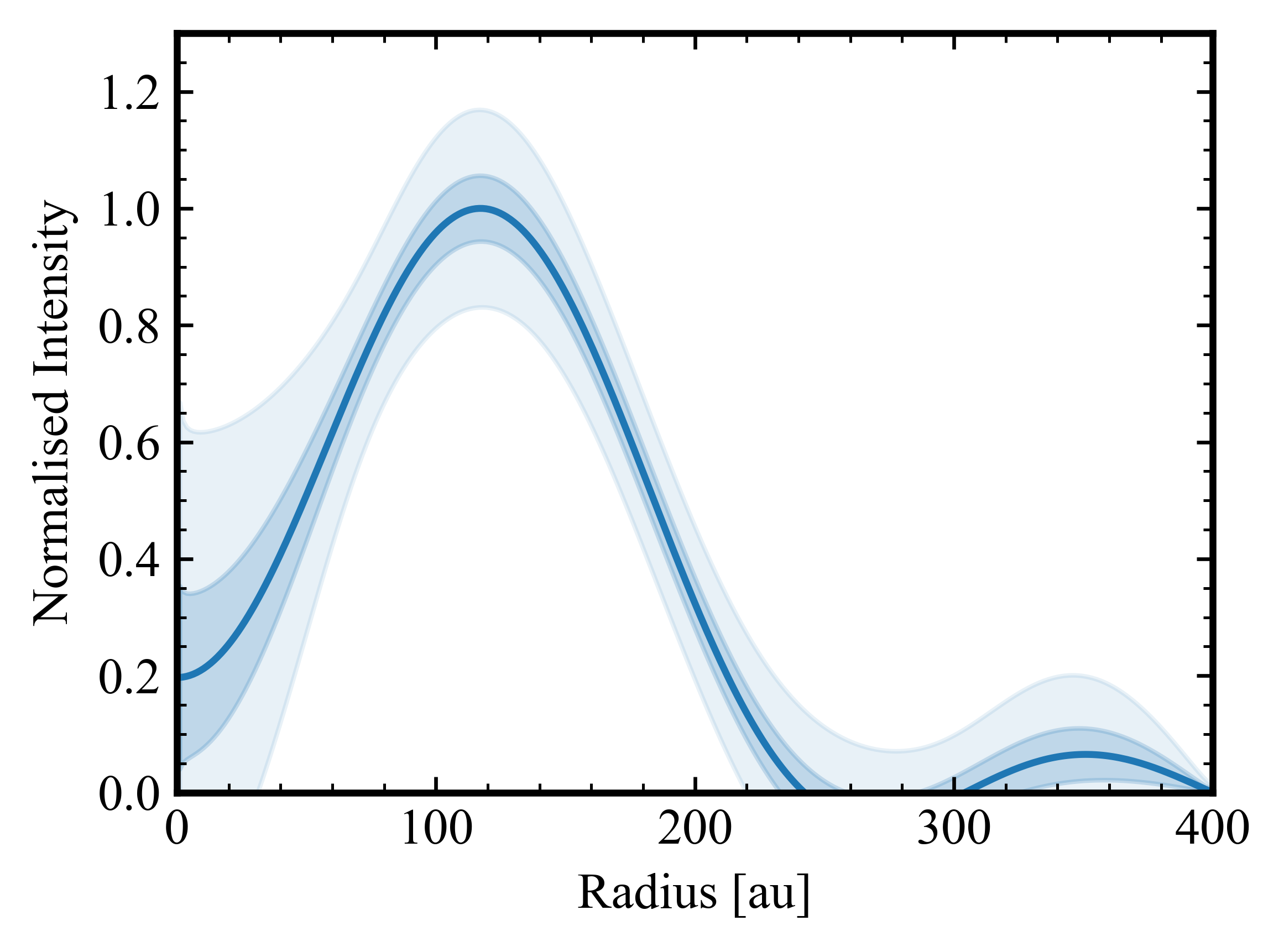}
    \caption{HD 161868}
    \label{fig:hd161868}
\end{subfigure}%
\begin{subfigure}{.33\textwidth}
\centering
    \includegraphics[width=\linewidth, trim = {0 0 0 0}, clip]{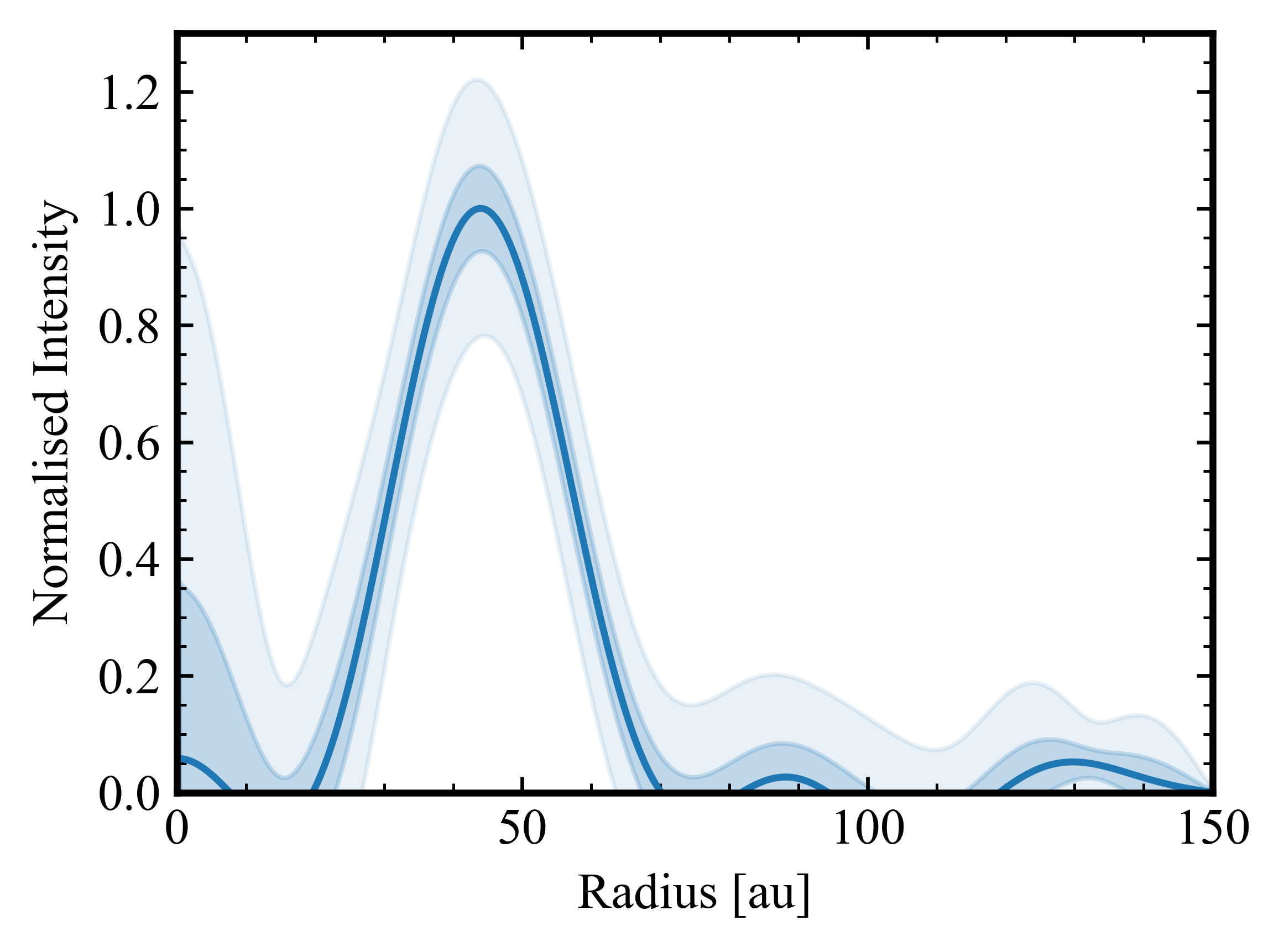}
    \caption{HD 191089}
    \label{fig:hd191089}
\end{subfigure}%
\begin{subfigure}{.33\textwidth}
\centering
\includegraphics[width=\linewidth,  trim = {0 0 0 0}, clip]{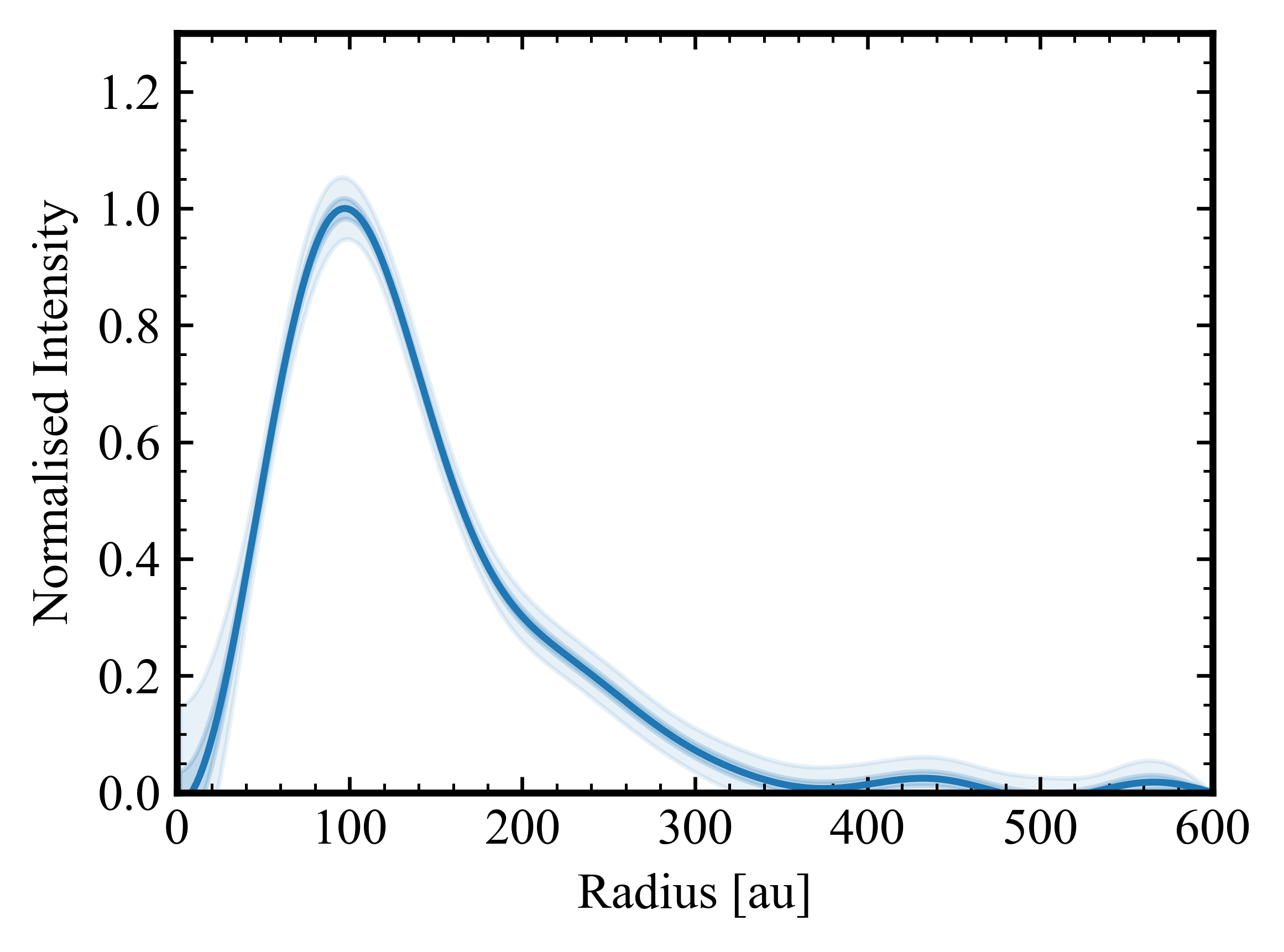}
    \caption{HD 9672}
    \label{fig:hd9672}
\end{subfigure}
\begin{subfigure}{.33\textwidth}
\centering
    \includegraphics[width=\linewidth, trim = {0 0 0 0}, clip]{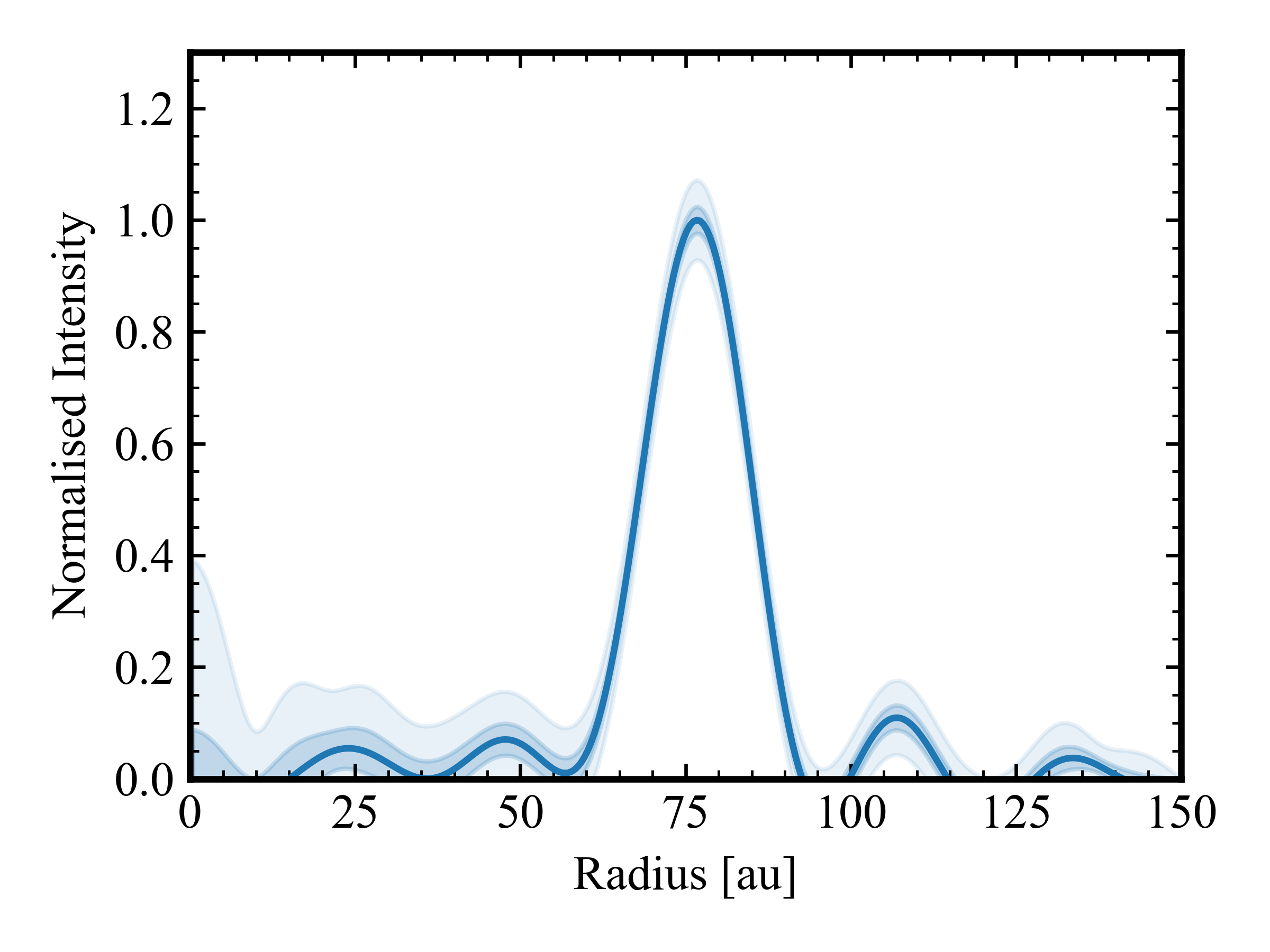}
    \caption{HD 109573}
    \label{fig:hd109573}
\end{subfigure}
\begin{subfigure}{.33\textwidth}
\centering
    \includegraphics[width=\linewidth, trim = {0 0 0 0}, clip]{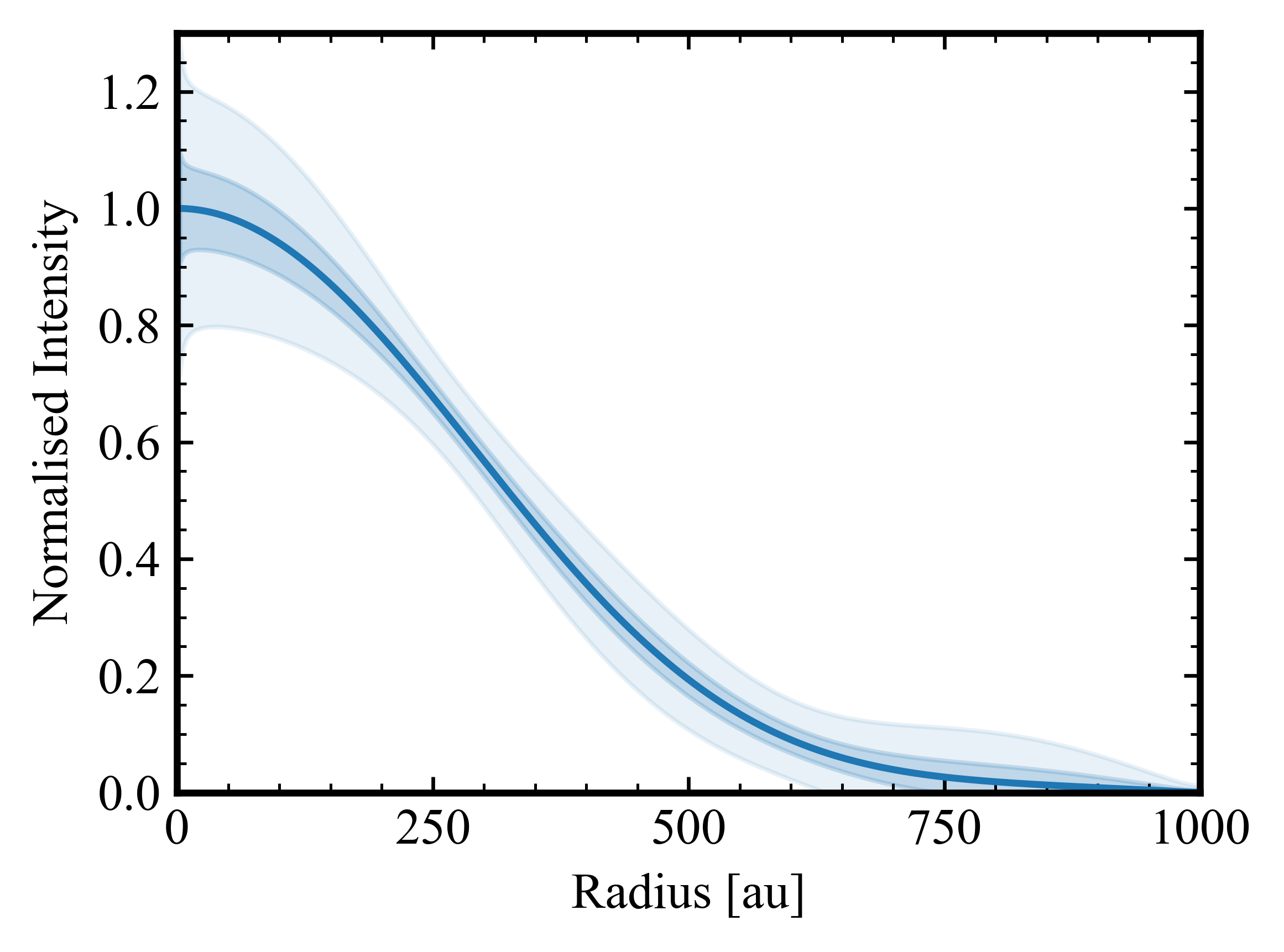}
    \caption{HD 158352}
    \label{fig:hd158352}
\end{subfigure}%
\begin{subfigure}{.33\textwidth}
\centering
    \includegraphics[width=\linewidth, trim = {0 0 0 0}, clip]{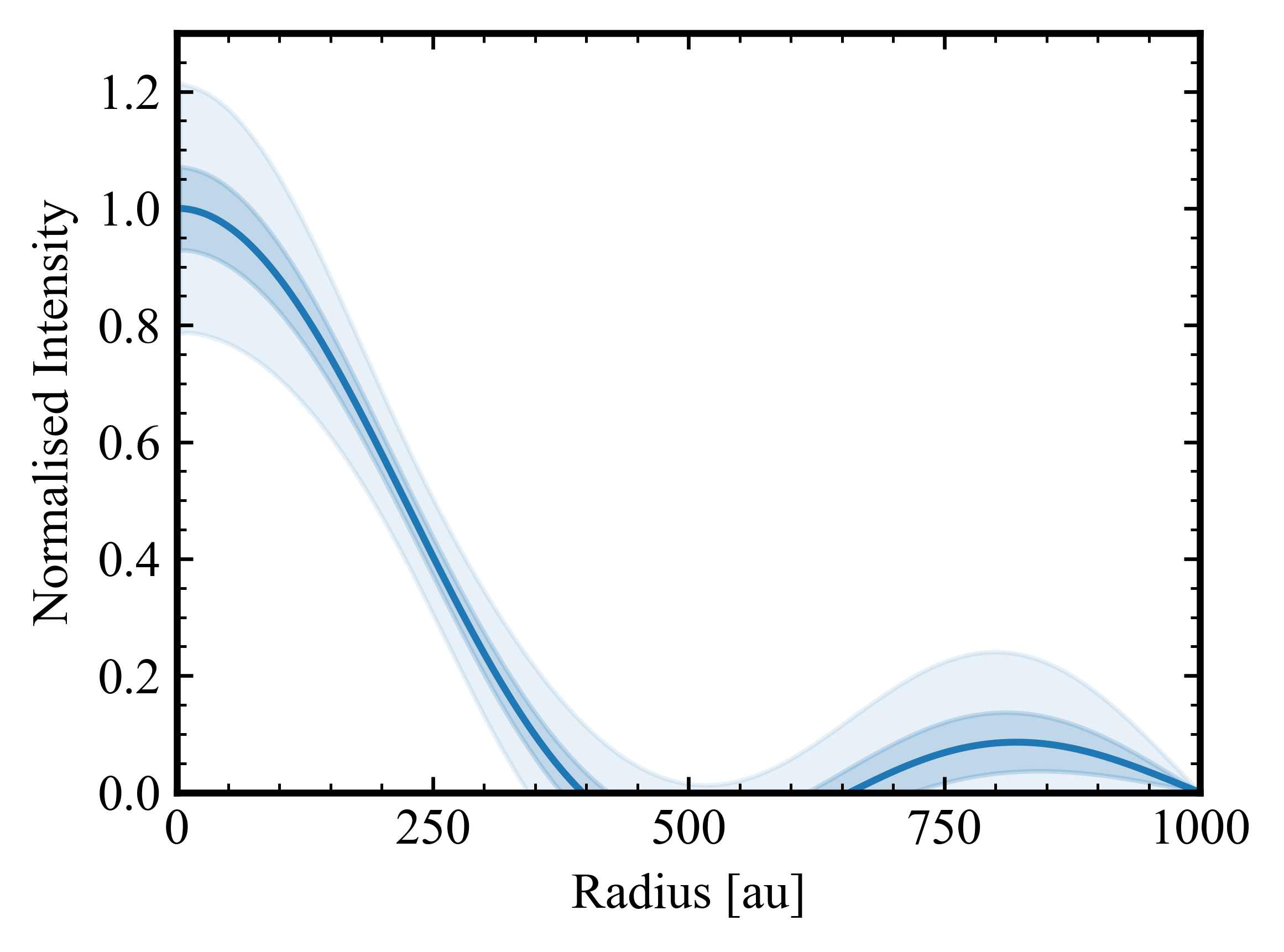}
    \caption{HD 50571}
    \label{fig:hd50571}
\end{subfigure}
\caption{Recovered intensity radial profiles for real debris discs; 1-3$\sigma$ confidence intervals are shown as shaded regions.}
\label{fig:real_radial_profiles}
\end{figure*}

\textsc{frank} is able to deproject the emission of a variety of edge-on debris discs. The recovered radial profiles in Figure \ref{fig:real_radial_profiles} reveal a range of features, including gaps and halos (discussed in \S\ref{subsubsec:gaps} and \S\ref{subsubsec:halos} respectively).

\subsubsection{Gap substructures}\label{subsubsec:gaps}
\textbf{AU Mic}\\
Figure \ref{fig:aumic} presents the radial profile for AU Mic recovered by \textsc{frank}. The majority of AU Mic's emission originates from a ring ${\sim}25$ au wide, centered at ${\sim}30$au. A small second peak in the intensity is found at ${\sim}10$ au, tentatively suggesting a second disc component and an intermediate gap. This morphology is consistent with parametric modelling done by \citet{Daley2019} and \citet{Marino2021}. Non-parametric modelling using \textsc{Rave} \citep{Han2022} also shows a similar feature. \newline

\noindent \textbf{HD 15115} \\
\noindent \citet{MacGregor2019} fit a parametric model to constrain the radial structure of HD~15115 and find evidence for a gap located at $58.9\pm4.5$ au of width $13.8\pm5.6$ au, consistent with the radial profile recovered here. However, the radial profile we obtain with \frank{} has a $\sim$3$\sigma$ negative region just interior to the disc. When we force \frank{} to a non-negative solution, the gap disappears, and it must therefore be treated with caution. \newline


\noindent \textbf{HD 92945} \\
Figure \ref{fig:hd92945} shows that the radial structure of HD~92945 features a gap centred at ${\sim}79$~au, with peaks at $\sim56$ au and ${\sim}100$ au, and an outer edge near $150$~au. This is in good agreement with \citet{Marino2019} and \citet{Marino2021}, who find evidence for a gap at $\sim73\pm3$ au and estimate the outer edge to be at ${\sim}140$ au. \newline

\noindent \textbf{HD 61005} \\
Figure \ref{fig:hd61005} presents the deprojected radial profile for HD~61005, with a prominent peak at ${\sim}68$~au and a secondary peak at  ${\sim}115$~au, which is stable in response variations of  $w_{\textrm{smooth}}$ and $\alpha$. Parametric modelling by \citet{MacGregor2018} shows evidence of a halo for HD~61005,  assuming an outer region of decaying surface density. The parametric model finds the peak of the distribution at 67~au, in agreement with the peak recovered here. However, the second peak in the \frank{} fit differs from the radially decaying power law that characterises a halo.

\subsubsection{Halo substructures}\label{subsubsec:halos}
\textbf{HD 10647} \\
The recovered radial profile for HD~10647 (q$^{1}$~Eri), presented in Figure \ref{fig:hd10647}, has a peak at ${\sim}85$~au, followed by a fit consistent with a wide decaying region, i.e., a halo (previously suggested by \citet{Lovell2021}). The small oscillations about this decaying shape are likely artifacts as found in one of our tests in \S\ref{subsec:radial_profiles_of_sim}. \newline

\noindent \textbf{HD 9672} \\
Figure \ref{fig:hd9672} shows HD~9672's (49~Ceti) radial profile with a peak at ${\sim}100$ au followed by a halo as the brightness profile decays out to ${\sim}320$ au. This is consistent with previous analysis by \citet{Hughes2017} and \citet{Higuchi2019}.

\subsection{HD 110058 - Analysis of a disc with an uncertain inclination}\label{subsec:hd110058}

\begin{figure}
    \centering
    \includegraphics[width=\linewidth, trim = {0 0 0 0}, clip]{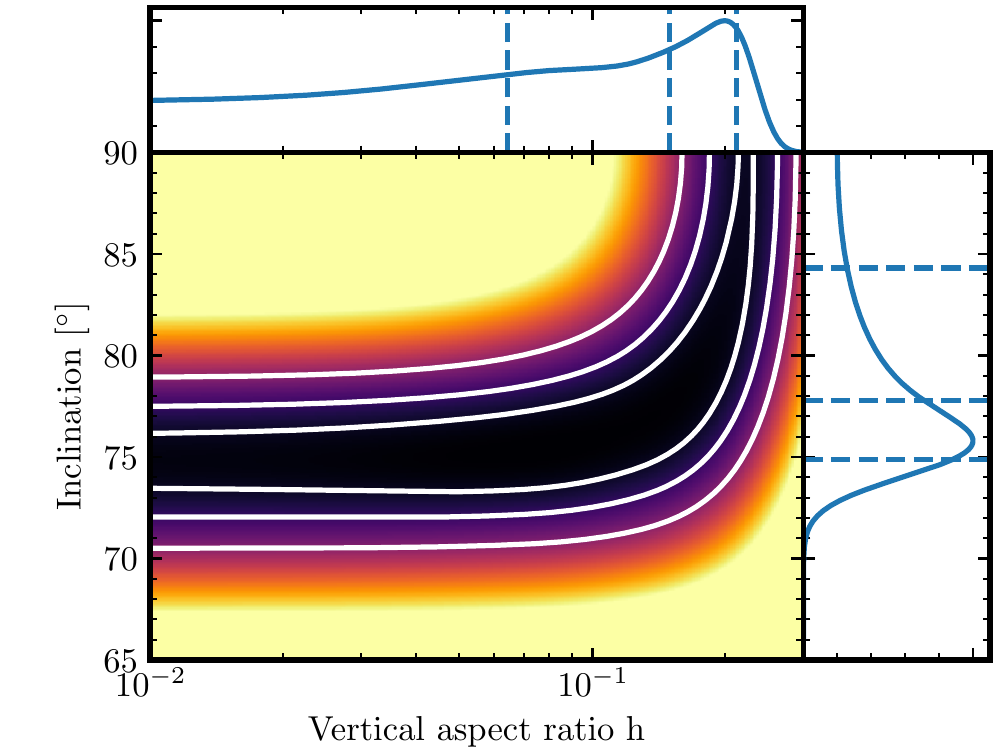}
    \caption{HD~110058's probability distribution as a function of the disc inclination and $h$, with darker colours representing the best fit. The contours represent the 68, 95 and 99.7\% confidence regions. The top and right panels show the marginalised probabilities. The vertical and horizontal dashed lines represent the 16th, 50th and 84th percentiles.}
    \label{fig:hd110058_chisq_contour}
\end{figure}

\begin{figure}
\centering
    \includegraphics[width=\linewidth, trim = {0 0 0 0}, clip]{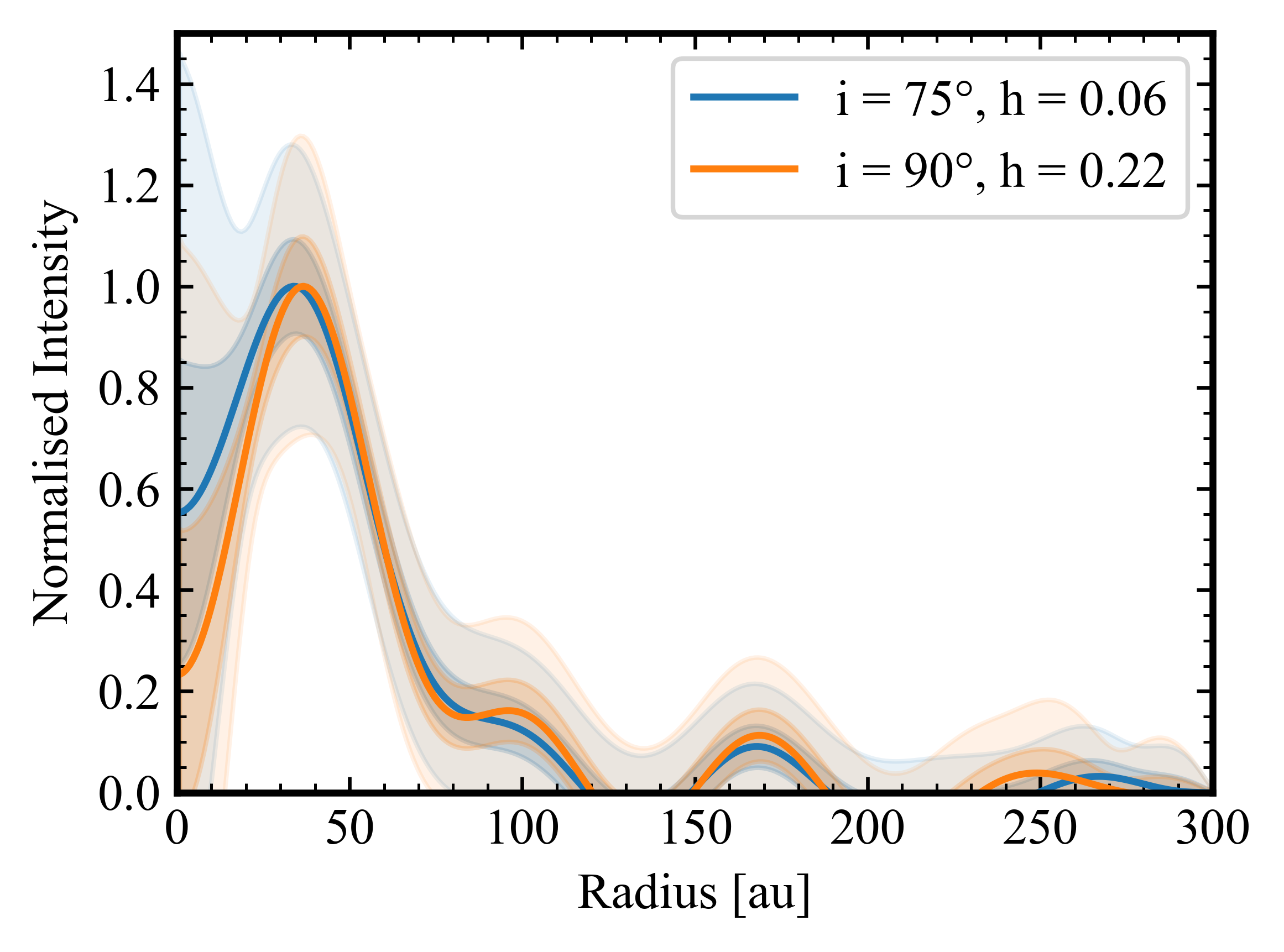}
\caption{The deprojected radial profile of HD~110058 for two viable inclination/aspect ratio combinations: $i = 75^{\circ}, h = 0.06$ and $i = 90^{\circ}, h = 0.22$. The $1-3 \sigma$ uncertainty regions are shaded.}
\label{fig:hd110058}
\end{figure}

The inclination of HD~110058 is not known precisely and has been estimated to be $\gtrsim80^{\circ}$ based on near-infrared observations \citep[]{Kasper2015, Esposito2020}. Recent ALMA observations resolved CO gas emission and determined a disc inclination that was likely to be $>80^{\circ}$ \citep{Hales2022}. As shown in \S\ref{subsubsec:wrong_inc}, assuming an incorrect inclination can yield an erroneous estimate of the aspect ratio. Therefore, we chose an agnostic approach and aim to fit the inclination and the aspect ratio jointly. To constrain both parameters simultaneously, we map the 2D probability distribution $P(h, $i$|\un{V}, \beta)$. The deprojection algorithm is run and the probability calculated for a range of $h$ and $i$. Figure \ref{fig:hd110058_chisq_contour} shows the 2D probability distribution with contours representing the 68, 95 and 99.7\% confidence regions. The 2D map reveals the degeneracy between these two parameters \citep[similar to what was found by][]{Hales2022}, as tests of simulated data demonstrated. The maximum is found at $\langle i \rangle = 75^{\circ}$, $\langle h \rangle = 0.06$, but this is only marginally better than other values with a wide range of inclinations up to 90$^{\circ}$ and aspect ratios from 0-0.3. The top and right panels show the marginalised probability distributions, which constrain the disc inclination to $i={77.8^{\circ}}^{+6.5}_{-2.9}$ and $h=0.15^{+0.06}_{-0.09}$ (68\% confidence). If we impose $i>80^{\circ}$ \citep[i.e. consistent with the scattered light observations,][]{Kasper2015, Esposito2020} we find $h=0.21\pm0.03$. These findings are consistent with \cite{Hales2022} that constrained $i$ and $h$ using a parametric model and the same data set.

 In order to compare the profiles produced at either end of the range of viable inclination and aspect ratios, we extract the radial profile assuming $h=0.06$ and $i=75^{\circ}$ (the values that maximize the posterior probability), and $h=0.22$ and $i=90^{\circ}$. The radial profiles are very similar, displaying large uncertainties as shown in Figure \ref{fig:hd110058}. Both fits show a clear peak at 35 au, with consistent decay. The main difference between the two profiles is at $r=0$, where the estimated intensity is larger for the $i=75^{\circ}$ solution.

\section{Discussion}
\label{sec:dis}

\subsection{Constraining the mass of a stirring body}\label{subsec:stirring_body}
The estimate of the aspect ratio offers an insight into the spread in inclination of the orbits of dust particles in the disc, $i_{\rm rms}=\sqrt{2} h$ \citep{Matra2019betapic}. The inclination dispersion is directly related to the relative velocities of the debris, yielding $v_{\rm rel}  \propto v_{\rm Kep}(r)h$, where $v_{\rm Kep}(r)$ is the Keplerian circular velocity at a distance $r$ from the star. Assuming this inclination dispersion arises from massive bodies embedded in the disc (self-stirring), we can use $h$ to constrain their size or mass. This is because a planet embedded in the disc will stir and excite the planetesimals'/debris' velocities up to at most its escape velocity and hence more massive bodies cause greater dynamical excitation \citep{Safronov1972, Goldreich2004, Schlichting2014}. 

Therefore, we estimate a rough lower limit on the diameter and mass ($D_{\rm stir}$ and $M_{\rm stir}$) of the stirring bodies by equating its escape velocity ($\sqrt{2\upi G D^2 \rho /3}$, where $\rho$ is the bulk density) to the relative velocity of particles ($\sqrt{1.25e_{\rm rms}^2+ i_{\rm rms}^2}v_\mathrm{Kep}$). This gives \citep[assuming $e_{\rm rms}=2i_{\rm rms}$,][]{Marino2021}\footnote{There is a typo in Equation 7 in \cite{Marino2021}. The exponents of $r$, $\rho$ and $M_{\star}$ should be -1/2, -1/2 and 1/2, respectively.}
\begin{align}
 \begin{split}
   D_{\rm stir}  =  590\ \mathrm{km} \left(\frac{h}{0.03} \right) \left(\frac{r}{100\ \mathrm{au}} \right)^{-1/2} \left(\frac{\rho}{2\ \mathrm{g\ cm^{3}}} \right)^{-1/2}  \\
   \left(\frac{M_\star}{1\ M_\odot} \right)^{1/2} 
 \end{split}
 \\
 \begin{split}
    M_{\rm stir}  = 3.5\times10^{-5}\ M_{\oplus} \left(\frac{h}{0.03} \right)^3 \left(\frac{r}{100\ \mathrm{au}} \right)^{-3/2} \left(\frac{\rho}{2\ \mathrm{g\ cm^{3}}} \right)^{-3/2}\\
    \left(\frac{M_\star}{1\ M_\odot} \right)^{3/2},
\end{split}
\end{align}
where $M_{\star}$ is the stellar mass. Table  \ref{tab:stirring_body_results} presents the derived minimum size of the stirring bodies using the values of $h$ that we found in \S\ref{sec:fitdata} and the systems' parameters.

\begin{table}
\centering
\captionof{table}{Estimates of the Minimum Mass of a Stirring Body. The stellar mass estimates come from: (1) \citet{Daley2019}, (2) \citet{Kervella2022}, (3) \citet{Marmier2013}, (4) \citet{MacGregor2019}, (5) \citet{Cataldi2020}, (6) \citet{Esposito2018}, (7) \citet{Desidera2015}, (8) \citet{Hughes2017}, (9) \citet{Hales2022}.  }
\begin{tabular}{l|l|l|l}
Disc     & \(\textup{M}_\star\) (\(\textup{M}_\odot\)) & $M_{\textrm{stir}}$ (\(\textup{M}_\oplus\))&$D_{\textrm{stir}}$ (km) \\ \hline
AU Mic   & 0.50 (1)  & $1.8\times10^{-5}$ & 480 \\
HD 10647 & 1.1 (3)  & $8.2\times10^{-5}$ & 780 \\
HD 15115 & 1.4 (4) & $2.4\times10^{-4}$ & 1100 \\
HD 32297 & 1.6 (5) & $1.3\times10^{-3}$ & 2000\\
HD 35841 & 1.3 (6)  & $1.8\times10^{-2}$ &4700\\
HD 61005 & 0.98 (2)  & $1.2\times10^{-4}$ & 900 \\
HD 92945 & 0.86 (7)  & $9.4\times10^{-5}$ & 820 \\
HD 161868 & 2.4 (2) & $1.3\times10^{-2}$ & 4200\\
HD 191089 & 1.3 (2)  & $7.5\times10^{-3}$ & 3500\\
HD 109573 & 2.2 (2)  &$9.0\times10^{-4}$& 1700\\ 
HD 9672  & 2.0 (8) &$4.3\times10^{-4}$ & 1400\\ 
HD 110058 & 1.8 (9) &$1.7\times10^{-1}$ & 10000\\
HD 158352 & 2.0 (2)& $1.4\times10^{-2}$ & 4300\\
HD 50571 & 1.4 (2) &$3.6\times10^{-3}$& 2800\\
\end{tabular}
\label{tab:stirring_body_results}
\end{table}

We find that if discs are self-stirred, the bodies stirring the disc should be at least $\sim500$~km in diameter, and 10 out of the 14 discs require bodies with diameters above 1000~km. HD~110058 would require large bodies with a size similar to Mars. The presence of such large bodies stirring the disc and resupplying the dust levels is challenging. This is because if we take the dust masses of these discs (typically in the range $0.01-0.5~M_{\oplus}$) and extrapolate these to such large sizes with standard size distributions, we find unrealistically high disc masses \citep{Krivov2021}. This tension could be solved by a very steep initial size distribution such that most of the disc mass is in bodies smaller than these dwarf-planets. Alternatively, the estimated disc thicknesses could arise from planet disc interactions, e.g. via planet-disc misalignment \citep{Wyatt1999} or via scattering \citep{Nesvorny2015}, in which case the discs would not need such large planetesimals and high masses.   




\subsection{Constraining the flaring index} \label{sec:flaring}

So far we have assumed that the vertical aspect ratio $h$ does not vary as a function of radius. For narrow discs this should not be an issue, but for wide discs $h$ could vary significantly between the disc inner and outer edges depending on what stirs the disc. For example, if the vertical stirring is due to secular interactions with a misaligned planet, after a few secular timescales the dispersion of inclinations (and $h$) will become constant as a function of semi-major axis (radius) and roughly equal to the original misalignment \citep[e.g.][]{Dawson2011}. However, if the vertical stirring was due to massive planetesimals embedded in the disc \citep[self-stirring,][]{Krivov2018stirring} or the secular timescale due to a misaligned planet was longer than the age of the system, the dispersion of inclinations and $h$ could vary significantly as a function of radius.  

\begin{figure}
    \centering
    \includegraphics[width=\linewidth, trim = {0 0 0 0}, clip]{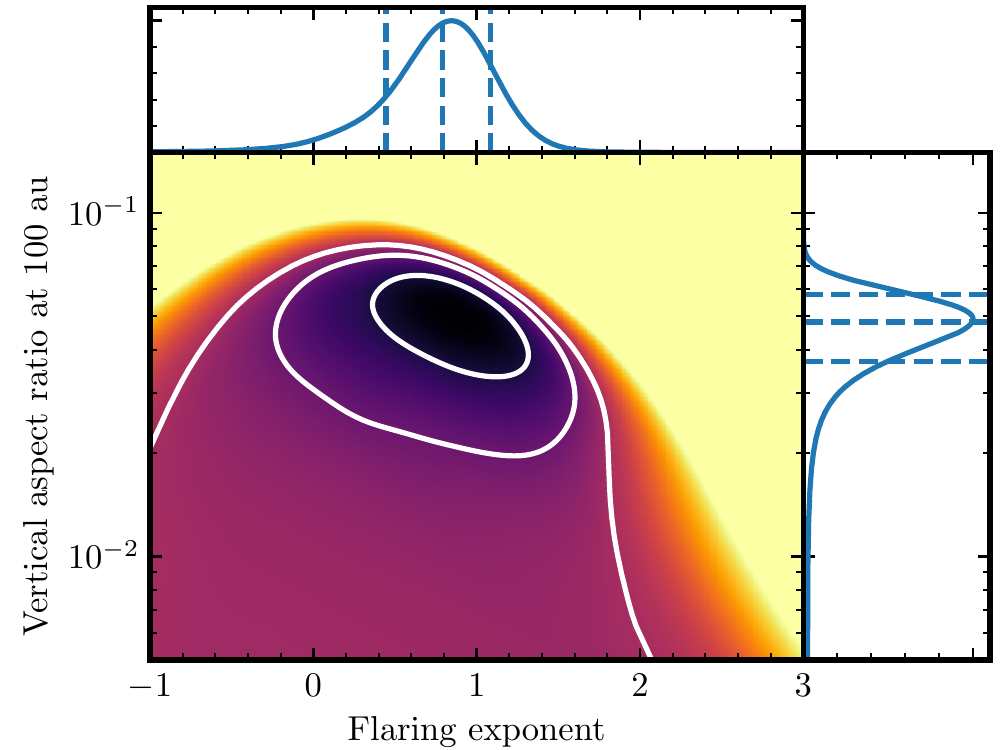}
    \caption{HD~9672's probability distribution as a function of the disc inclination and the flaring index with darker colours representing the best fit. The contours represent the 68, 95 and 99.7\% confidence regions. The top and right panels show the marginalised probabilities. The vertical and horizontal dashed lines represent the 16th, 50th and 84th percentiles.}
    \label{fig:hd9672_chisq_contour}
\end{figure}

Of the studied discs, HD~9672 (49~Ceti) is the most promising to study whether $h$ could vary as a function of radius since it has a wide disc that is very well resolved and detected with a very high signal to noise. In order to constrain the flaring index, we adapt the scale height definition to 
\begin{equation}
    h(r)=h_{100\ \mathrm{au}} (r/100\ \mathrm{au} )^{\gamma-1},
\end{equation}
where $h_{100\ \mathrm{au}}$ is the aspect ratio at 100~au, and $\gamma$ is the flaring index, which we have assumed it is equal to 1 so far. Values lower than 1 indicate an aspect ratio that decreases with radius, whereas values larger than 1 correspond to aspect ratios increasing with radius. We proceed to map the 2D posterior probability distribution as a function of $h_{100\ \mathrm{au}}$ and $\gamma$. Figure~\ref{fig:hd9672_chisq_contour} shows the posterior probability distribution, which has a maximum at $h_{100\ \mathrm{au}}=0.05$ and $\gamma=0.9$. The marginalised probabilities constrain $h_{100\ \mathrm{au}} = 0.048^{+0.010}_{-0.011}$ and  $\gamma = 0.79^{+0.29}_{-0.35}$. This means that HD~9672's observations are in good agreement with a constant $h$ ($\gamma=1$), but we cannot rule out cases in which $h$ increases by a factor ${\sim}2$ or decreases by a factor ${\sim}4$ between the disc extent from 100-300~au. Exploring in detail the flaring index of each one of the discs in our sample is beyond the scope of this paper and would likely require higher-resolution observations.

\subsection{Comparison with previous estimates of $h$}
\label{subsec:compare_to_old_h}

In order to assess the validity of our estimates of $h$, we compare them with the results in the literature obtained by fitting a range of parametric models to 5 of the discs in the studied sample. These are AU~Mic, HD~10647 (q$^1$~Eri), HD~92945, HD~109573 (HR~4796) and HD~110058. Overall, we find a good agreement with both estimates differing by less than $3\sigma$. For AU~Mic, \cite{Marino2021} found $h=0.0021\pm0.004$ and we found $0.002\pm0.002$. For HD~10647, \cite{Lovell2021} found $h=0.048\pm0.004$ and we found $0.037^{+0.008}_{-0.007}$. For HD~92945, \cite{Marino2021} found $h=0.061\pm0.020$ and we found $h=0.04\pm0.01$. For HD109573, \cite{Kennedy2018} found $h=0.038\pm0.005$ and we found $h=0.052\pm0.003$. For HD~110058, \cite{Hales2022} found $h=0.214\pm0.024$ and we found $0.21\pm0.03$ using the same prior ($i>80^{\circ}$). We can also compare our estimate of $h$ for AU~Mic with the non-parametric estimate by \cite{Han2022} using \textsc{Rave}, and we also find a good agreement. Assuming a similar inclination of $88.5^{\circ}$, they derive an average height of 0.8~au, which at the disc peak radius of $\sim30$~au translates to $h=0.026$. This estimate is consistent with ours.

Finally, to assess if the derived uncertainties are reasonable we compare them with the literature values quoted above and also with those derived by Matr\`a et al. in prep that fitted a parametric model for all the discs studied here. The latter study fitted a disc model where the radial and vertical density distribution of dust follows a Gaussian distribution, i.e. simpler models compared to our non-parametric radial fits. This procedure and model are the same as in \cite{Marino2016}. We find that our derived uncertainties are consistent with the ones from the parametric model fits, with ours being only 13\% smaller on average. The slightly smaller uncertainties are likely due to our approach of using a fixed inclination (except for HD~110058) as the inclination and $h$ can become degenerate. The uncertainties in inclination from parametric fits tend to be small ($1^{\circ}$ on average), which explains why the difference is only 13\% on average. Therefore, we conclude that our uncertainties are (to first order) well estimated.

\subsection{Emissivity with a Gaussian distribution} \label{sec:gaussian}

Throughout this paper we have assumed discs have an emissivity that is approximately Gaussian as a function of height. This requires:
\begin{itemize}
    \item The dust vertical distribution is Gaussian with scale height $H(r)$. This would be the case if orbital inclinations have a Rayleigh distribution as expected for ensembles of interacting planetesimals and solids \citep{Ida1992}. However, there are scenarios where multiple dynamical populations co-exist at the same radius creating more complex distributions. This is the case of the Kuiper belt and $\beta$~Pic's disc \citep{Brown2001, Matra2019betapic}. In such a case, a single value of $h$ is not appropriate and our derived values could be dominated by the most excited of the two populations \citep[see model comparisons in ][]{Matra2019betapic}. Assessing the multiple populations scenario requires high-resolution observations that can resolve the disc height. This is not possible in the observations presented here where $H$ is marginally resolved.
    \item $H(r)$ does not vary strongly within the grain size range contributing the most to the disc emission at a single wavelength, which at millimetre wavelengths corresponds to roughly grain sizes of $0.1-10\times$ the wavelength (see Appendix~\ref{sec:emissivity}). Using collisional models that considered viscous stirring and collisional damping, \cite{Pan2012} showed that the velocity dispersion may vary with size in a collisional cascade, leading to significant variations of $H$ within the relevant size range. Recent observations of AU~Mic's debris disc support this possibility showing a tentative increase in the vertical height with wavelengths between 0.45mm and 1.3mm \citep{Vizgan2022}; however, the increase with wavelength is inconsistent with standard collisional models. Moreover, recent observations of HD~16743 show an almost identical scale height at NIR and millimitre wavelengths \citep{Marshall2023}. If instead, stirring is dominated by external perturbers (e.g. misaligned or eccentric planet) $H$ might behave differently and possibly remain independent of size. Figure~\ref{fig:vertical_dist} shows the emissivity\footnote{Opacity calculations are presented in Appendix~\ref{sec:emissivity}. Note that here we have assumed that the dust temperature is independent of size. However, small grains tend to be hotter which could increase slightly their total contribution to the emission at mm wavelengths.} as a function of $z$ for four different cases where $H$ or the inclination dispersion ($i_{\rm rms}$) could be independent of grain size ($a$) or vary as $a^{p}$ as proposed by \cite{Pan2012}. For $p\gtrsim0.3$ the departure from a Gaussian (blue line) is significant and thus our assumption would not be valid anymore. It is worth noting that a non-Gaussian emissivity was inferred for $\beta$~Pic and interpreted as multiple dynamical populations \citep{Matra2019betapic}, however a single dynamical population with a size dependent $i_{\rm rms}$ might explain the observations as well.   
    \item The dust temperature does not vary with height. Since debris discs are very optically thin the equilibrium temperature of grains is not a function of height above the midplane.
\end{itemize}

\begin{figure}
    \centering
    \includegraphics[width=1.0\columnwidth]{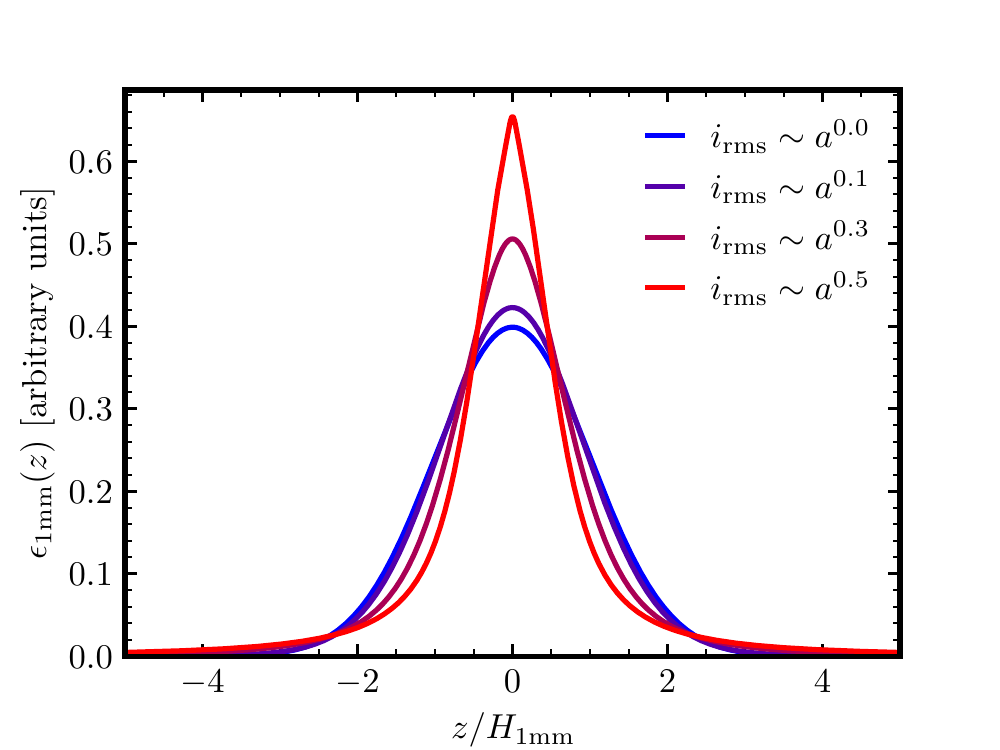}
    \caption{Emissivity at 1~mm as a function of height for dust grains with a size distribution $N(a){\sim}a^{-3.5}$ from 1~$\mu$m to 10~cm and with $H$ or inclination dispersion ($i_{\rm rms}$) proportional to $a^{p}$. The integrated emissivity has been normalized to an arbitrary constant and the height $z$ has been normalized to the scale height of 1~mm sized grains. The dust temperature is assumed to be independent of grain size.}
    \label{fig:vertical_dist}
\end{figure}

\section{Conclusions and summary}
\label{sec:conclusions}

In this paper, we have presented a new approach to simultaneously deproject the emission of optically thin and axisymmetric circumstellar discs observed by ALMA or any interferometer (even if edge-on) and constrain their vertical structure. Given their low optical depth, this is particularly useful for debris disc studies.  We first show how the deprojected visibilities of an optically thin edge-on disc are not different from a face-on disc. Therefore, methods such as \textsc{Frankenstein} \citep{Jennings2020} that can retrieve the radial intensity profiles of discs directly from the observed visibilities, can also be used to deproject the emission of edge-on discs as long as they are axisymmetric and optically thin along the line-of-sight.

Furthermore, we show the effect the disc scale height, $H(r)$, has on the visibilities and how this effect can be incorporated into \frank{} assuming the vertical distribution is Gaussian. We develop a new extension to \frank{} in which $H(r)$ is an input, and using simulated observations we show how the model can accurately retrieve the radial profile if $H(r)$ is known.  More importantly, $H(r)$ is usually unknown and directly linked to the dispersion of orbital inclinations, hence it is a key property to constrain the level of dynamical excitation. Therefore, we demonstrate how sampling the posterior distribution with a grid allows us to constrain $H(r)$ if it is parameterised (e.g. assuming $h=H/r$ is constant). Our simulations show the algorithm can  constrain the vertical thickness even in extreme cases where the \texttt{CLEAN} beam is only a tenth of $H$, as long as the observations have a high signal-to-noise and the disc inclination is well known. This new approach to retrieve the radial profile and constrain the vertical thickness of discs with \textsc{frank} provides two major benefits versus parametric models. Firstly, no functional form is assumed for the radial profile. Secondly, \textsc{frank} can produce a constraint for both the vertical and radial structure of disc in minutes, far faster than the hours typically taken by MCMC methods, due to their need to sample several parameters and Fourier transform a model several times.

We applied this new method to 16 highly inclined debris discs observed by ALMA and successfully constrained the aspect ratio, $h$, for discs with both known and uncertain inclinations (sampling the posterior distribution with 1 ($h$) or 2 free parameters ($h$ and $i$ - Table \ref{tab:real_h_results} and \S\ref{subsec:hd110058}). The aspect ratios of the debris discs in our sample range from $0.020\pm0.002$ to $0.22\pm0.03$ (for AU~Mic and HD~110058 respectively). We find a tentative correlation between the aspect ratio and fractional width of discs (Figure \ref{fig:real_disc_sum} bottom panel), indicating a possible bimodal distribution where discs with large fractional widths tend to have larger $h$ values and vice versa. If true this could mean that disc stirring could be responsible for the large widths of some discs. This new extension to \textsc{frank} also allows us to constrain how the aspect ratio might vary as a function of the disc's radius as expected in some dynamical scenarios. We apply this to HD~9672 (49~Ceti), a wide disc with a very high signal-to-noise ALMA observation, and find a result consistent with $h$ being constant with radius (although the flaring index is still highly uncertain). We also compare our results with the $h$ estimates derived from parametric models applied to the same data, finding a good agreement in both the estimates and derived uncertainties.

Assuming that the discs are self-stirred, the values of $h$ we derive require planetesimals with masses of at least $2\times10^{-5}\ M_{\oplus}$ and diameters of at least 500~km. Such large planetesimals would imply unphysically large disc masses unless the size distribution was very steep and the disc mass dominated by smaller planetesimals. Alternatively, the discs could be stirred by planets via scattering or secular interactions.  

Finally, the deprojected radial profiles reveal a range of structures at a higher resolution than previous images. These include halos (i.e. smoothly decreasing surface brightness) around HD~10647 (q$^1$~Eri) and HD~9672 (49~Ceti), and gaps around HD~15115, HD~92945 and HD~61005. The latter had not been inferred before (see subfigure \ref{fig:hd61005}) due to the comparatively lower resolution of \texttt{CLEAN} images. Instead, the second peak that we found was previously interpreted as a halo. This emphasizes the benefits of using \textsc{frank}.

\section*{Acknowledgements}

We would like to thank Luca Matr\`a for providing the calibrated and reduced ALMA data for most of the discs we analysed. Throughout this project, Sebastian Marino was supported by a Junior Research Fellowship from Jesus College, University of Cambridge, and currently by a Royal Society University Research Fellowship. Richard Booth is supported by a Royal Society University Research Fellowship. This paper makes use of the following ALMA data: ADS/JAO.ALMA\#2012.1.00198.S, ADS/JAO.ALMA\#2015.1.00032.S, ADS/JAO.ALMA\#2015.1.00633.S, ADS/JAO.ALMA\#2015.1.01260.S, ADS/JAO.ALMA\#2016.1.00104.S, ADS/JAO.ALMA\#2016.1.00880.S, ADS/JAO.ALMA\#2017.1.00167.S, ADS/JAO.ALMA\#2017.1.00200.S, ADS/JAO.ALMA\#2017.1.00467.S, ADS/JAO.ALMA\#2018.1.00500.S,
ADS/JAO.ALMA\#2019.1.01517.S. ALMA is a partnership of ESO (representing its member states), NSF (USA), and NINS (Japan), together with NRC (Canada), MOST and ASIAA (Taiwan), and KASI (Republic of Korea), in cooperation with the Republic of Chile. The Joint ALMA Observatory is operated by ESO,
AUI/NRAO, and NAOJ.

\section*{Data Availability}

 The data underlying this article will be shared on reasonable request to the corresponding author. The ALMA data are publicly available and can be queried and downloaded directly from the ALMA archive at https://almascience.nrao.edu/asax/. Frankenstein is publicly available and the version used to take into account the disc vertical thickness can be found at https://github.com/discsim/frank.


\bibliographystyle{mnras}
\bibliography{refs} 




\appendix

\section{Statistical estimation of the disc scale height}
\label{app:metrics}

\begin{figure}
    \centering
    \includegraphics[width=\columnwidth]{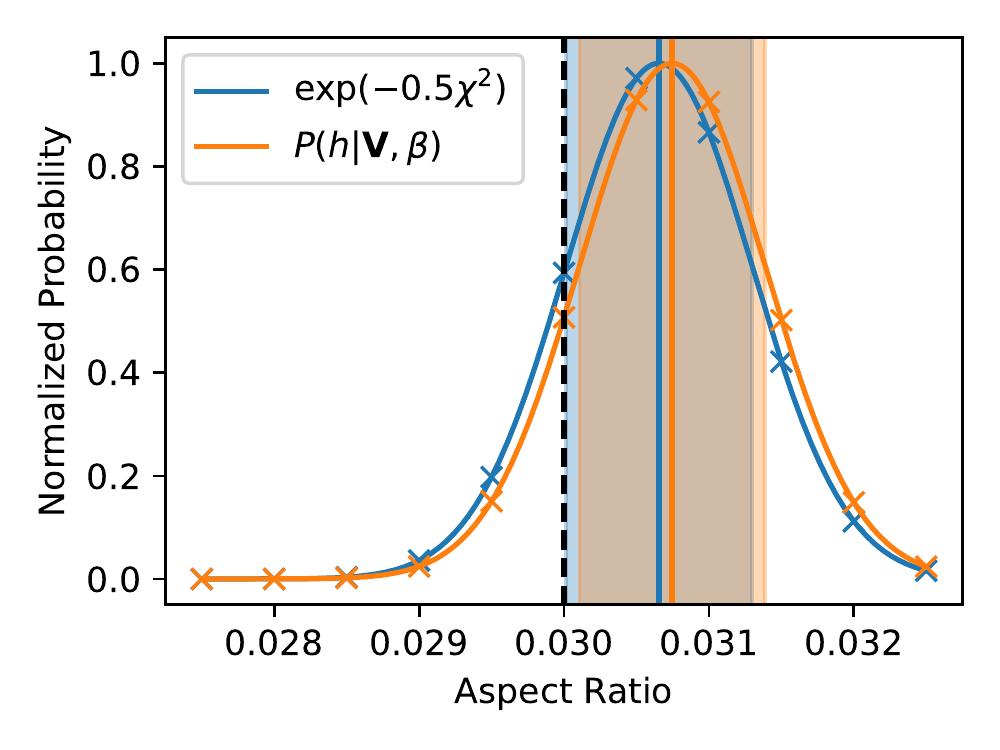}
    \caption{A comparison of two different metrics for determining the best fitting disc aspect ratio applied to the double ring test problem. The vertical lines show the best-fitting value and the coloured bands the 1-$\sigma$ confidence interval around the best fit. The black-dashed line is the ground truth value. Both metrics are similar, producing best fit values that agree to within 1-$\sigma$.}
    \label{fig:Chi2_vs_Bayes}
\end{figure}

In this paper we have used the (Laplace-approximated) Bayesian evidence, $P(h|\un{V}, \beta)$, to determine the disc scale height. A comparison between the estimates produced via the Bayesian evidence and previous estimates of the scale height derived from parametric fits to the same data (\S\ref{subsec:compare_to_old_h}) show good agreement. Here we argue that the inferred values of the disc scale height are unlikely to be sensitive to the precise details of the metric used. An alternative, frequentist, approach to determining the best fit is to use the $\chi^2$ metric to estimate the disc scale height. I.e.,
\begin{equation}
    \chi^2 = \sum_i\frac{\left[V_i - V_s(\un{I},h)_i\right]^2}{\sigma_i^2} = - 2 \log P(\un{V}|\un{I},h)  + \mathrm{const.}
\end{equation}
where $w_i = 1/\sigma_i^2$ and last equality follows because the noise on the visibility data is approximately Gaussian. Finding the parameters (brightness profile, scale height) that minimize the $\chi^2$ therefore is equivalent to finding the model for which the data is most probable. 

We show a comparison between (Laplace-approximated) Bayesian evidence and $\exp(-\chi^2/2)$ in \autoref{fig:Chi2_vs_Bayes}, from which it is clear that a frequentist and Bayesian interpretation should lead to similar inferences. The explanation as to why the two metrics agree closely is simply that the variation in the Bayesian evidence with $h$ is dominated by the change in the $\chi^2$.

\section{Grain sizes that dominate the emission} \label{sec:emissivity}

\begin{figure}
\centering
\includegraphics[width=\columnwidth]{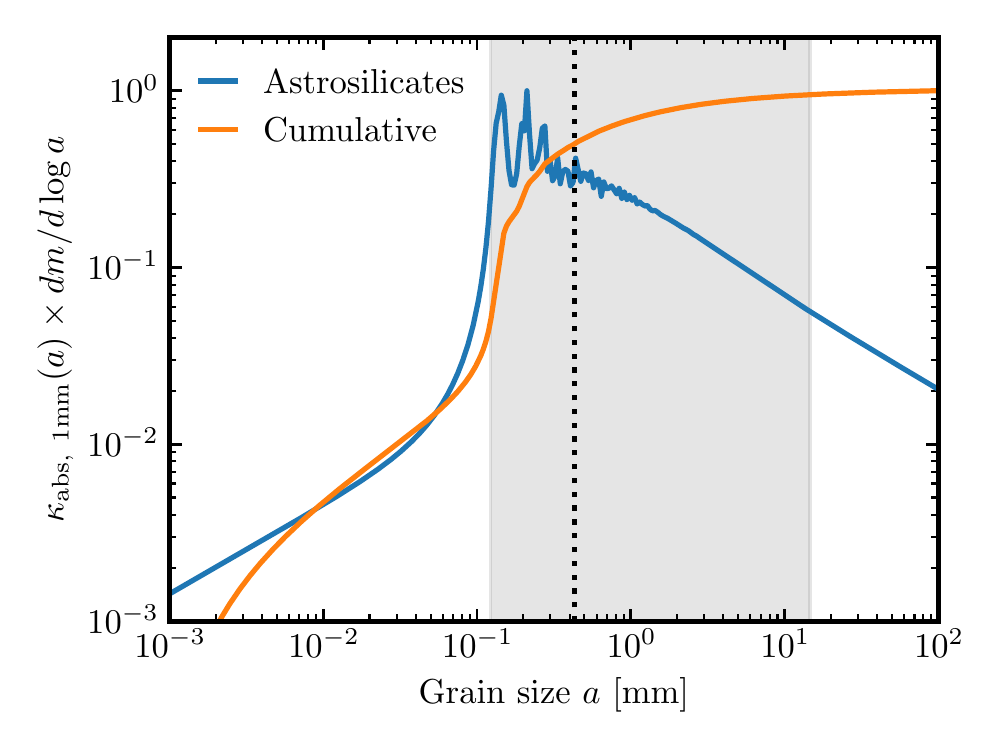}
\caption{Absorption opacity at 1~mm as a function of grain size and weighted by the mass distribution with logarithmic size bins (blue line in arbitrary units). The orange curve shows the cumulative absorption cross-section. The grey shaded region and vertical dotted line represent the grain sizes contributing 90\% of the emission and the median, respectively. }
\label{fig:emissivity}    
\end{figure}

Here we want to estimate the contribution of different grains sizes to the emission at a given wavelength \citep[e.g. as done by][]{Wyatt2002}. To do this we assume a size distribution from 1~$\mu$m up to 10~cm with an exponent of -3.5 and focus on a wavelength of 1~mm (representative of the wavelengths  studied here). We use Mie Theory to compute the absorption opacity \citep{BohrenHuffman1983} and we assume grains have an astrosilicate composition \citep{Draine2003}. Figure~\ref{fig:emissivity} shows the absorption opacity at 1~mm as a function of grain size and weighted by the mass distribution with logarithmic size bins, i.e. the contribution from each grain size to the total absorbing and emitting area as a proxy for the emission (assuming temperatures do not vary significantly for grains larger than ${\sim}0.1$~mm). We find that 90\% of the emission (grey area) arises from grains sizes from 0.12 to 15 times the wavelength. This highlights how the emission at ALMA wavelengths is dominated by grain sizes spanning roughly 2 orders of magnitude and centered at the wavelength. 

\bsp	
\label{lastpage}
\end{document}